\documentclass[12pt]{article}
\usepackage{amsmath,epsfig,graphicx,verbatim,subfigure} 
\include{epsf} 
\def\ltap{\raisebox{-.6ex}{\rlap{$\,\sim\,$}} \raisebox{.4ex}{$\,<\,$}} 
\def\gtap{\raisebox{-.6ex}{\rlap{$\,\sim\,$}} \raisebox{.4ex}{$\,>\,$}}

\newcommand\as{\alpha_{\mathrm{S}}} 
\newcommand\f[2]{\frac{#1}{#2}} 
\def\beq{\begin{equation}} 
\def\eeq{\end{equation}} 
\def\beeq{\begin{eqnarray}} 
\def\eeeq{\end{eqnarray}} 
 
\def\to{\rightarrow}
 
\def\nn{\nonumber}

\def\msbar{{\overline {\rm MS}}} 
\def\qt{q_T}

\begin{document} 
\begin{titlepage}
\begin{flushright}
ZU-TH 17/11
\end{flushright}
\renewcommand{\thefootnote}{\fnsymbol{footnote}}
\vspace*{1cm}

\begin{center}
{\Large \bf Transverse-momentum resummation:\\[0.4cm]
Higgs boson production at the Tevatron and the LHC} \\
\vskip 0.2cm
\end{center}

\par \vspace{2mm}
\begin{center}
\vspace{.5cm}
{\bf Daniel de Florian}$^{(a)}$, 
{\bf Giancarlo Ferrera}$^{(b,c)}$,\\ {\bf Massimiliano Grazzini}$^{(d)}$\footnote{On leave of absence from INFN, Sezione di Firenze, Sesto Fiorentino, Florence, Italy.}~~and~~{\bf Damiano Tommasini}$^{(b,d)}$\\

\vspace{0.8cm}

${}^{(a)}$
Departamento de F\'\i sica, FCEYN, Universidad de Buenos Aires,\\
(1428) Pabell\'on 1 Ciudad Universitaria, Capital Federal, 
Argentina\\\vskip .05cm
${}^{(b)}$
Dipartimento di Fisica e Astronomia, Universit\`a di Firenze and\\
INFN, Sezione di Firenze,
I-50019 Sesto Fiorentino, Florence, Italy\\\vskip .05cm
${}^{(c)}$
Dipartimento di Fisica, Universit\`a di Milano and\\
INFN, Sezione di Milano, I-20133 Milan, Italy\\
${}^{(d)}$Institut f\"ur Theoretische Physik, Universit\"at Z\"urich, CH-8057 
Z\"urich, Switzerland
\vspace{1cm}
\end{center}

\par \vspace{2mm}
\begin{center} {\large \bf Abstract} \end{center}
\begin{quote}
\pretolerance 10000

We consider the transverse-momentum ($q_T$)
distribution 
of Standard Model Higgs bosons  produced by gluon fusion
in hadron collisions. 
At small $q_T$ ($q_T\ll m_H$, $m_H$ being the mass of the Higgs boson),
we resum the logarithmically-enhanced contributions due
to multiple soft-gluon emission to all order in QCD perturbation theory.
At intermediate and large values of $q_T$ ($q_T\ltap m_H$), 
we consistently combine resummation with the known 
fixed-order results.
%
We use the most advanced  perturbative information
that is available at present:
next-to-next-to-leading logarithmic resummation combined with the
next-to-leading fixed-order calculation.
We extend previous results including exactly  
 all the perturbative terms up to order $\as^4$ in our computation and, 
after integration over $q_T$, we recover
the known next-to-next-to-leading order result for the 
total cross section.
We present numerical results at the Tevatron and the LHC,
together with an estimate
of the corresponding uncertainties.
Our calculation is implemented in an updated version of the numerical code
{\ttfamily HqT}.
\end{quote}

\vspace*{\fill}
\begin{flushleft}
September 2011
\end{flushleft}
\end{titlepage}

\setcounter{footnote}{1}
\renewcommand{\thefootnote}{\fnsymbol{footnote}}
\section{Introduction}

One of the major tasks of the physics program at high-energy hadron colliders,
 such as the Fermilab Tevatron and the CERN LHC, is the
search for the Higgs boson and the study of its properties. 

Gluon--gluon fusion, through a heavy-quark (mainly top-quark) loop,
is the main production mechanism of the Standard Model (SM) Higgs boson over the 
entire range of Higgs boson masses (100 GeV $\ltap m_H \ltap 1$ TeV) to be investigated at the LHC.
At the Tevatron the gluon fusion process, followed by the decay  
$H \rightarrow WW \rightarrow l^+l^-\nu \bar\nu$, gives the dominant
contribution to the Higgs signal in the range of mass 
$140$~GeV $\ltap m_H \ltap 180$~GeV. In this mass region, 
first constraints beyond the LEP lower bound of $114.4$~GeV \cite{Barate:2003sz}
 were established:
the SM Higgs boson was excluded at 95\% confidence level
by CDF and D0 collaborations  
in the mass range $156$ GeV $<m_H<177$ GeV \cite{CDF:2011cb}.
The first results of the ATLAS and CMS collaborations presented at EPS 2011 conference \cite{eps11}, and updated
for Lepton Photon 2011 \cite{lp11}, dramatically
extend the excluded region over most of the mass range between 145 and 466 GeV.

The above exclusion relies on accurate theoretical predictions \cite{Anastasiou:2008tj,deFlorian:2009hc}
for the
inclusive $gg\to H$ cross section, which is now known up to next-to-next-to-leading order (NNLO) \cite{NNLOtotal},
with the inclusion of soft-gluon contributions up to next-to-next-to-leading logarithmic accuracy (NNLL) \cite{Catani:2003zt}, and two-loop electroweak effects \cite{ew}
\footnote{Updated predictions for the inclusive Higgs production cross sections at the LHC are presented in Ref.~\cite{Dittmaier:2011ti}.}.

In this paper we consider the
transverse momentum ($q_T$) spectrum of the SM Higgs boson $H$
produced by the gluon fusion mechanism.
This observable is of direct importance in the experimental search.
A good knowledge of the $q_T$ spectrum can help to set up strategies
to improve the statistical significance.
When studying the $q_T$ distribution 
of the Higgs boson in QCD perturbation theory it is convenient to define
two different regions of $q_T$.
In the large-$q_T$ region ($q_T\sim m_H$), where the transverse momentum is
of the order of the Higgs boson mass $m_H$,
perturbative QCD
calculations based on the truncation of the perturbative series at a
fixed order in $\as$ 
are theoretically justified.
In this region, the QCD radiative corrections are known up to the next-to-leading
order (NLO) \cite{deFlorian:1999zd,Ravindran:2002dc,Glosser:2002gm}
and QCD corrections beyond the NLO are evaluated in Ref. \cite{deFlorian:2005rr},
by implementing threshold 
resummation at the next-to-leading logarithmic (NLL) level.

In the small-$q_T$ region ($q_T\ll m_H$),
where the bulk of the events is produced, 
the convergence of the fixed-order expansion is spoiled by the presence of 
large logarithmic terms, $\as^n\ln^m (m^2_H/q_T^2)$.
To obtain reliable predictions, 
these logarithmically-enhanced terms 
have to be systematically
resummed to all
perturbative orders \cite{Dokshitzer:hw,Catani:vd,Catani:2000vq,Bozzi:2005wk,Catani:2010pd}.
It is then important to consistently match the resummed and fixed-order
calculations 
at intermediate values of $q_T$, in order
to obtain accurate QCD predictions for the entire range of transverse momenta.

The resummation of the logarithmically enhanced terms is effectively (approximately) performed
by standard Monte Carlo event generators.
In particular, MC@NLO \cite{Frixione:2002ik} and POWEG \cite{Nason:2004rx} combine soft-gluon resummation through the parton shower
with the leading order (LO) result valid at large $q_T$, thus achieving a result with formal NLO accuracy.

The numerical program {\ttfamily HqT} \cite{Bozzi:2005wk}
implements
soft-gluon resummation up to NNLL
accuracy \cite{deFlorian:2000pr} combined with fixed-order perturbation theory up to NLO in the large-$q_T$ region \cite{Glosser:2002gm}.
The program is used by the Tevatron and LHC experimental collaborations
to reweight the $q_T$ spectrum of the Monte Carlo event generators used in the analysis and
is thus of direct relevance in the Higgs boson search.

The program {\ttfamily HqT} is based on the transverse-momentum resummation formalism described
in Refs. \cite{Catani:2000vq, Bozzi:2005wk, Catani:2010pd}, 
which is valid for 
a generic process in which
a high-mass system of non strongly-interacting particles is produced 
in hadron--hadron collisions.
The method has so far been applied to the production of
the SM Higgs boson \cite{Bozzi:2005wk,Bozzi:2003jy,Bozzi:2007pn},
single vector bosons \cite{Bozzi:2008bb,Bozzi:2010xn},
$WW$ \cite{Grazzini:2005vw} and $ZZ$ \cite{Frederix:2008vb} pairs,
slepton pairs \cite{Bozzi:2006fw}, and
Drell-Yan lepton pairs in polarized collisions \cite{Jiro}.

In this paper we update and extend the phenomenological analysis presented 
in Ref. \cite{Bozzi:2005wk}. In particular, we implement the
exact value of the NNLO hard-collinear coefficients ${\cal H}_N^{H(2)}$ computed
in Ref. \cite{Catani:2007vq,Catani:2011kr},
and the recently derived 
value of the NNLL coefficient $A^{(3)}$ \cite{Becher:2010tm}.

We use the most advanced  perturbative information that is available at present: NNLL resummation  at small
$q_T$ and the fixed-order NLO calculation at large $q_T$.
We present  numerical results
for Higgs production
at the Tevatron Run II and at the LHC 
and we perform a detailed study of the
perturbative uncertainties. We also consider the normalized $q_T$ spectrum and discuss its
theoretical uncertainties.
Our calculation for the $q_T$ spectrum 
is implemented in the updated version of the numerical code {\ttfamily HqT}, 
which can be downloaded from \cite{hqt}.
Other phenomenological studies of the Higgs boson
$q_T$ distribution, which combine resummed and fixed-order perturbative
results at 
various levels of theoretical accuracy,
can be found in Refs.~\cite{Balazs:2000wv}--\cite{Mantry:2010mk}.

The paper is organized as follows. In Sect.~\ref{sec:theory} we briefly review 
the resummation formalism of Refs. \cite{Catani:2000vq,Bozzi:2005wk,Catani:2010pd} 
and its application
to Higgs boson production.
In Sect.~\ref{sec:results} 
we present numerical results
for Higgs boson production
at the Tevatron and the LHC.
In Sect.~\ref{sec:summa} we summarize our results.

\section{Transverse-momentum resummation}
\label{sec:theory}

In this section we briefly recall the 
main points of the transverse-momentum resummation approach proposed in  
Refs. \cite{Catani:2000vq,Bozzi:2005wk,Catani:2010pd}.
We consider the specific case of a Higgs boson $H$  produced
by gluon fusion. As recently pointed out in Ref. \cite{Catani:2010pd},
the gluon fusion $q_T$-resummation formula has a different
structure than the resummation formula for $q \bar q$ 
annihilation. The difference originates from the 
collinear correlations that are a specific feature 
of the perturbative evolution of colliding
hadron into gluon partonic initial states.
These gluon collinear correlations produce, in the small-$q_T$ region,
coherent spin correlations between the helicity
states of the initial-state gluons and definite azimuthal-angle correlations
 between the final-states particles of the observed high-mass system.
Both these kinds of correlations have no analogue for $q \bar q$ annihilation
processes in the small-$q_T$ region.
In the case of Higgs boson production, being $H$ a spin-$0$ scalar particle,
the azimuthal correlations vanishes and only gluon spin correlations are
present \cite{Catani:2010pd}.
 
We consider the inclusive hard-scattering process
\begin{equation}
h_1(p_1) + h_2(p_2) \;\to\; H (m_H,q_T ) + X ,   
\label{first}
\end{equation}
where $h_1$ and $h_2$ are the colliding hadrons with momenta
$p_1$ and $p_2$, 
$m_H$ and $q_T$  are  the Higgs boson
mass and transverse momentum respectively,
and $X$ is an arbitrary and undetected final state. 

According to the QCD factorization theorem
the corresponding transverse-momentum 
differential cross 
section $d\sigma_H/dq_T^2$ can be written as
\begin{equation}
\label{dcross}
\f{d\sigma_H}{d q_T^2}(q_T,m_H,s)= \sum_{a,b}
\int_0^1 dx_1 \,\int_0^1 dx_2 \,f_{a/h_1}(x_1,\mu_F^2)
\,f_{b/h_2}(x_2,\mu_F^2) \;
\f{d{\hat \sigma}_{H,ab}}{d q_T^2}(q_T, m_H,{\hat s};
\as(\mu_R^2),\mu_R^2,\mu_F^2) 
\;\;,
\end{equation}
where $f_{a/h}(x,\mu_F^2)$ ($a=q,{\bar q}, g$)
are the parton densities of the colliding hadron $h$ 
at the factorization scale $\mu_F$, 
$d\hat\sigma_{H,ab}/d{q_T^2}$ are the perturbative QCD 
partonic cross sections, 
$s$ ($\hat s = x_1 x_2 s$) 
is the square of the 
hadronic (partonic) centre--of--mass  energy, 
and $\mu_R$ is the renormalization 
scale\,\footnote{Throughout the paper we use parton densities 
$f(x,\mu_F^2)$
and running
coupling $\as(\mu_R^2)$ 
as defined in the $\msbar$ scheme.}. 

In the region where 
$q_T \sim  m_H$,
the QCD perturbative
series is controlled by a small expansion parameter, 
$\as(m_H)$,
and fixed-order calculations are
theoretically justified. In this region, 
the QCD radiative corrections are known up to
NLO \cite{deFlorian:1999zd,Ravindran:2002dc,Glosser:2002gm}. 
In the small-$q_T$ region 
($q_T\ll m_H$),
the convergence of the fixed-order
perturbative expansion is spoiled
by the presence 
of powers of large logarithmic terms, 
$\as^n\ln^m (m_H^2/q_T^2)$ (with $1\leq m \leq 2n-1)$.
To obtain reliable predictions these terms have to be resummed to all orders.

We perform the resummation
at the level of the partonic cross section, which
is decomposed~as
\begin{equation}
\label{resplusfin}
\f{d{\hat \sigma}_{H,ab}}{dq_T^2}=
\f{d{\hat \sigma}_{H,ab}^{(\rm res.)}}{dq_T^2}
+\f{d{\hat \sigma}_{H,ab}^{(\rm fin.)}}{dq_T^2}\; .
\end{equation}
The first term on the right-hand side
contains all the logarithmically-enhanced contributions, at small $q_T$,
and has to be evaluated to all orders in $\as$.
The second 
term 
is free of such contributions and
can thus be computed at fixed order in perturbation theory. 
To correctly take into account the kinematic constraints of
transverse-momentum conservation, the resummation procedure has to be carried out
in the impact parameter space $b$. 
Using the Bessel transformation between the conjugate variables 
$q_T$ and  $b$,
the resummed component $d{\hat \sigma}^{({\rm res.})}_{H,ac}$
can be expressed as
\begin{equation}
\label{resum}
\f{d{\hat \sigma}_{H,ac}^{(\rm res.)}}{dq_T^2}(q_T,m_H,{\hat s};
\as(\mu_R^2),\mu_R^2,\mu_F^2) 
= 
\int_0^\infty db \; \f{b}{2} \;J_0(b q_T) 
\;{\cal W}^H_{ac}(b,m_H,{\hat s};\as(\mu_R^2),\mu_R^2,\mu_F^2) \;,
\end{equation}
where $J_0(x)$ is the $0$th-order Bessel function.
The resummation structure of ${\cal W}^H_{ac}$ can 
be  organized in exponential  form
considering
the Mellin $N$-moments ${\cal W}^H_N$ of ${\cal W}^H$ with respect to the variable 
$z=m_H^2/{\hat s}$ at fixed 
$m_H$\,\footnote{For the sake of simplicity we are presenting the resummation
formulae only for 
the specific case of 
the diagonal terms in the flavour space. 
In general, the exponential
is replaced by an exponential matrix with respect
to the partonic indeces (a 
detailed discussion of the general case can be found in
Ref. \cite{Bozzi:2005wk}).},
\begin{align}
\label{wtilde}
{\cal W}^H_{N}(b,m_H;\as(\mu_R^2),\mu_R^2,\mu_F^2)
&={\cal H}_{N}^H\left(m_H, 
\as(\mu_R^2);m_H^2/\mu^2_R,m_H^2/\mu^2_F,m_H^2/Q^2
\right) \nonumber \\
&\times \exp\{{\cal G}_{N}(\as(\mu^2_R),L;m_H^2/\mu^2_R,m_H^2/Q^2
)\}
\;\;,
\end{align}
were we have defined the logarithmic expansion parameter 
$L\equiv \ln ({Q^2 b^2}/{b_0^2})$,
and $b_0=2e^{-\gamma_E}$ ($\gamma_E=0.5772...$ 
is the Euler number).

The scale $Q\sim m_H$, appearing in the right-hand side of Eq.~(\ref{wtilde}), 
named resummation scale \cite{Bozzi:2005wk}, 
parameterizes the
arbitrariness in the resummation procedure.
As a matter of fact the argument of the resummed logarithms 
can always be rescaled as
$\ln ({m_H^2 b^2})= \ln ({Q^2 b^2}) + \ln ({m_H^2/Q^2})$
(as long as $Q \sim m_H$ and independent of $b$).
Although ${\cal W}^H_{N}$ 
(i.e., the product
${\cal H}_{N}^H \times \exp\{{\cal G}_{N}\}$) does not depend on $Q$ when
evaluated to all perturbative orders, its explicit dependence on $Q$
appears when ${\cal W}^H_{N}$ is computed by truncation of the resummed
expression at some level of logarithmic accuracy (see Eq.~(\ref{exponent})
below). 
As in the case of $\mu_R$ and $\mu_F$, variations of 
$Q$ around $m_H$ can thus be used to estimate the
uncertainty from yet uncalculated 
logarithmic corrections at higher orders.

The  form factor $\exp\{ {\cal G}_N\}$ is 
{\itshape universal} (process independent)\,\footnote{It 
only  depends on the partonic channel that produces the Born cross section.
It is thus usually called quark  or gluon Sudakov form factor.}
and contains all
the terms $\as^nL^m$ with $1 \leq m \leq 2n$, 
that order-by-order in $\as$ are logarithmically divergent 
as $b \to \infty$ (or, equivalently, $q_T\to 0$). 
Furthermore, due to the {\itshape exponentiation} property,  
all the logarithmic contributions to ${\cal G}_N$ with
$n+2 \leq m \leq 2n$ are vanishing.
The  exponent ${\cal G}_N$ 
can  be systematically expanded as
\begin{align}
\label{exponent}
{\cal G}_{N}(\as, L;m_H^2/\mu^2_R,m_H^2/Q^2)&=L 
\;g^{(1)}(\as L)+g_N^{(2)}(\as L;m_H^2/\mu_R^2,m_H^2/Q^2)\nn\\
&+\f{\as}{\pi} g_N^{(3)}(\as L;m_H^2/\mu_R^2,m_H^2/Q^2)
+{\cal O}(\as^n L^{n-2})
\end{align}
where the term $L\, g^{(1)}$ resums the leading logarithmic (LL) 
contributions $\as^nL^{n+1}$, the function $g_N^{(2)}$ includes
the NLL contributions $\as^nL^{n}$ \cite{Catani:vd}, 
$g_N^{(3)}$ controls the NNLL 
terms $\as^nL^{n-1}$ \cite{deFlorian:2000pr,Becher:2010tm}
and so forth. The explicit form of the functions
$g^{(1)}$, $g_N^{(2)}$ and $g_N^{(3)}$ can be found in Ref. \cite{Bozzi:2005wk}.

The process {\itshape dependent} function ${\cal H}_N^H$ 
does not depend on the impact parameter $b$ and it 
includes all the perturbative
terms that behave as constants as 
$b \to \infty$. 
It can thus be expanded in powers of $\as=\as(\mu_R^2)$:
\begin{align}
\label{hexpan}
{\cal H}_N^H(m_H,\as;m_H^2/\mu^2_R,m_H^2/\mu^2_F,m_H^2/Q^2)&=
\sigma_H^{(0)}(\as,m_H)
\Bigl[ 1+ \f{\as}{\pi} \,{\cal H}_N^{H,(1)}(m_H^2/\mu^2_F,m_H^2/Q^2) 
\Bigr. \nn \\
&\!\!\!\!\!\!+ \Bigl.
\left(\f{\as}{\pi}\right)^2 
\,{\cal H}_N^{H,(2)}(m_H^2/\mu^2_R,m_H^2/\mu^2_F,m_H^2/Q^2)+
{\cal O}(\as^3)
\Bigr] \;\;,
\end{align}
where $\sigma_H^{(0)}(\as,m_H)$
is the partonic cross section at the Born level.
The
first order ${\cal H}_{N}^{H,(1)}$ \cite{Kauffman:cx}
and the second order ${\cal H}_{N}^{H,(2)}$ \cite{Catani:2007vq,Catani:2011kr}
coefficients  in Eq.~(\ref{hexpan}),
for the case
of Higgs boson production in the large-$M_t$ approximation, are known.

To reduce the impact of unjustified higher-order contributions in 
the large-$q_T$ region,
the logarithmic variable $L$ in Eq.~(\ref{wtilde}), 
which diverges for $b\to 0$, 
is actually replaced  by 
${\widetilde L}\equiv \ln \left({Q^2 b^2}/{b_0^2}+1\right)$ \cite{Bozzi:2005wk, Bozzi:2003jy}.
The variables $L$ and ${\widetilde L}$ are equivalent when $Qb\gg 1$ 
(i.e. at small values $q_T$), but they 
lead to a different behaviour
of the form factor at small values of $b$. 
An additional and relevant
consequence of this replacement 
is that, after inclusion of the finite component (see Eq.~(\ref{resfin})),  
we exactly recover the fixed-order perturbative value of the total cross section
upon integration of the $q_T$  distribution over $q_T$
(i.e., the contribution of the resummed terms
vanishes upon integration over $q_T$).

The finite component of the transverse-momentum cross section $d\sigma_H^{({\rm fin.})}$
(see Eq.~(\ref{resplusfin}))
 does not contain large logarithmic terms
in the small-$q_T$ region,
it can thus be evaluated by truncation of the perturbative series
at a given fixed order.
In practice it is computed as follows
\begin{equation}
\label{resfin}
\Bigl[ \f{d{\hat \sigma}_{H,ab}^{(\rm fin.)}}{d q_T^2} \Bigr]_{\rm f.o.} =
\Bigl[\f{d{\hat \sigma}_{H,ab}^{}}{d q_T^2}\Bigr]_{\rm f.o.}
- \Bigl[ \f{d{\hat \sigma}_{H,ab}^{(\rm res.)}}{d q_T^2}\Bigr]_{\rm f.o.} \;,
\end{equation} 
where we have introduced the subscript ${\rm f.o.}$  to denote the perturbative truncation of the
various terms.
This matching procedure 
 combines the resummed and the 
finite component of
the partonic cross section by avoiding double-counting in the intermediate $q_T$-region
and allows us 
to achieve a prediction with uniform theoretical accuracy 
over 
the entire range 
of transverse momenta.

In summary,
 to carry out the resummation at NLL+LO accuracy, we need the
inclusion of the functions $g^{(1)}$, $g_N^{(2)}$,
${\cal H}_N^{H,(1)}$, in Eqs.~(\ref{exponent},\ref{hexpan}),
together with the evaluation of the finite component at LO 
(i.e. at ${\cal O}(\as)$) in Eq.~(\ref{resfin});
the addition of the functions $g_N^{(3)}$ and ${\cal H}_N^{H,(2)}$, together 
with the finite component at NLO (i.e. at ${\cal O}(\as^2)$)
leads to the NNLL+NLO 
accuracy\,\footnote{The evaluation of the  second-order coefficient ${\cal H}_{N}^{H,(2)}$
for complex values of $N$,
necessary to perform the inverse Mellin transform,  is obtained using
the numerical results of Ref. \cite{Blumlein:2000hw}.}.
We point out that our best theoretical prediction 
(NNLL+NLO)   includes the {\em full} NNLO 
perturbative contribution in the small-$\qt$ region plus the NLO correction at large-$q_T$.
In particular,  the NNLO  result for the total cross section  
is exactly recovered upon integration
over $q_T$ of the differential cross section $d \sigma_H/dq_T$ at NNLL+NLO
accuracy.

Finally we recall 
that the resummed form factor 
$\exp \{{\cal G}_N(\as(\mu_R^2),{\widetilde L})\}$
has a singular behaviour, related to the presence of the Landau 
pole in the QCD running coupling, at 
the values of $b$ where $\as(\mu_R^2) {\widetilde L} \geq \pi/\beta_0$ 
($\beta_0$ is the first-order coefficient of the QCD $\beta$ function).
To perform 
the  inverse 
Bessel
transformation with respect to the impact parameter $b$ a prescription
is thus necessary.
We deal with this
singularity
by using the regularization prescription of
Refs. \cite{Laenen:2000de,Kulesza:2002rh}: 
the singularity is avoided by deforming the 
integration contour in the complex $b$ space.

\section{The $q_T$ spectrum of the Higgs boson at the Tevatron and the LHC}
\label{sec:results}

In this section 
we consider Higgs boson production by gluon fusion at the Tevatron ($\sqrt{s}=1.96$ TeV)
and the LHC ($\sqrt{s}=7$ TeV and $14$ TeV). 
We present our resummed results at NNLL+NLO accuracy,
and we compare them with the NLL+LO results. 
For the Tevatron we choose $m_H=165$ GeV.
For the LHC at $\sqrt{s}=7$ and $\sqrt{s}=14$ TeV we fix $m_H=165$ GeV and $m_H=125$ GeV, respectively.

The results we present in this section are obtained with  an updated version of the 
numerical code {\ttfamily HqT} \cite{hqt}. 
The new version of this code was improved with respect
to the one used in Ref. \cite{Bozzi:2005wk}. 
The main differences regard the implementation  
of the second-order coefficients  
${\cal H}_{N}^{H,(2)}$
computed in Ref. \cite{Catani:2007vq} 
(the numerical results
in Ref. \cite{Bozzi:2005wk} were obtained by using a reasonable approximation 
of this coefficient)
and the use of the recently derived 
value of the coefficient $A^{(3)}$ \cite{Becher:2010tm}  
which  contributes to the NNLL
function $g_N^{(3)}$ 
(the results in Ref. \cite{Bozzi:2005wk} were obtained by using the 
$A^{(3)}$  value from threshold resummation \cite{Moch:2004pa}). 
We have checked the quantitative effect
of the exact values of ${\cal H}^{H,(2)}$ and $A^{(3)}$  at the Tevatron and the LHC. 
We find that the effect is generally small (at the level of about $1-2\%$ at the LHC at $14$ TeV, $2-3\%$
at the Tevatron, and at the LHC with 7 TeV).
We also find that the exact values of ${\cal H}^{H,(2)}$ and $A^{(3)}$ have the same qualitative impact:
it makes the $q_T$-spectrum (slightly) harder.


The calculation is performed strictly in the large-$M_t$ approximation.

The hadronic $q_T$ cross section at NNLL+NLO (NLL+LO) accuracy
is computed
by using NNLO (NLO) parton distributions functions (PDFs)
with $\as(\mu_R^2)$ evaluated at 3-loop (2-loop) order.
This choice of the order of the parton densities and $\as$
is fully justified both in the small-$q_T$ region
(where the calculation of the partonic cross section includes the complete
NNLO (NLO) result and is controlled by NNLL (NLL) 
resummation) and in the intermediate-$q_T$ region
(where the calculation is  
constrained by the value of the NNLO (NLO) total cross section).
Recent sets of parton densities, which are obtained by analyses of various collaborations, are
presented in Refs. \cite{othpdf,Martin:2009iq,Alekhin:2009ni,JimenezDelgado:2009tv,Ball:2011uy}.
Since the main purpose of our work is the study of the $q_T$ distribution up to the NNLL+NLO,
we consider here only the PDFs sets
of Refs.~\cite{Martin:2009iq,Alekhin:2009ni,JimenezDelgado:2009tv,Ball:2011uy},
which provide NNLO parton densities with $N_f=5$ (effectively) massless quarks.
Moreover, to avoid multiple presentations of similar results, we use the MSTW2008 parton densities
unless otherwise stated (the results in Ref. \cite{Bozzi:2005wk} were obtained by using the MRST2004 
set \cite{Martin:2004ir}).

As discussed in Sect.~\ref{sec:theory}, the 
resummed calculation depends on the factorization and 
renormalization scales and on the resummation scale $Q$. 
Our convention to compute factorization 
and renormalization scale uncertainties is to consider
independent variations of $\mu_F$ and $\mu_R$ by a factor of two around 
the central values $\mu_F=\mu_R=m_H$
(i.e. we consider the range $m_H/2\leq \{\mu_F,\mu_R\}\leq 2\,m_H$), with the constraint
$0.5 \leq \mu_F/\mu_R \leq 2$. 
Similarly, we follow Ref. \cite{Bozzi:2005wk} and
choose $Q=m_H/2$ as central value of the resummation scale,
considering scale variations in the 
range $m_H/4 < Q < m_H$.

\begin{figure}[htp]
\begin{center}
\subfigure[]{\label{fig1a}
\includegraphics[width=0.47\textwidth]{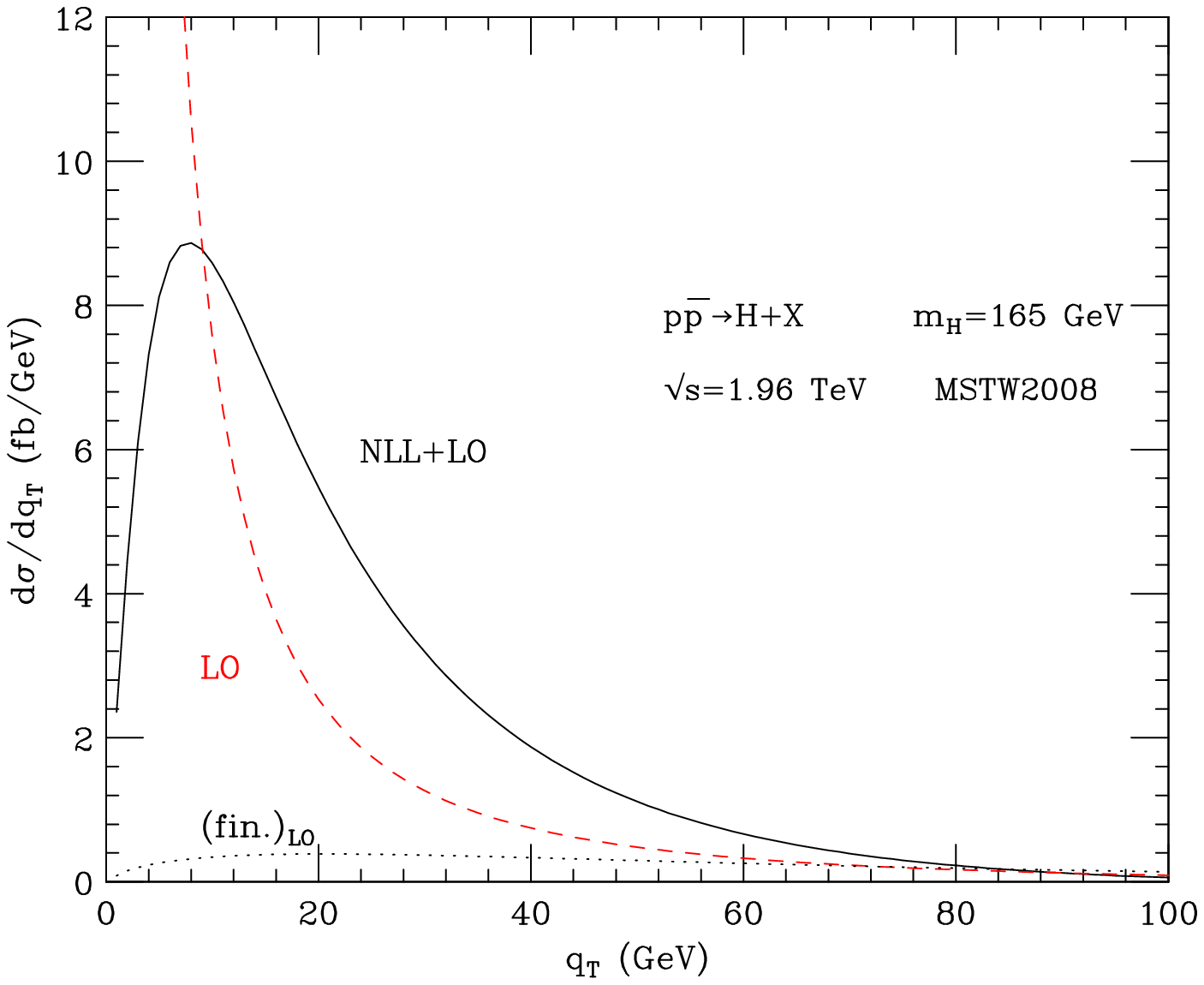}}
\subfigure[]{\label{fig1b}
\includegraphics[width=0.47\textwidth]{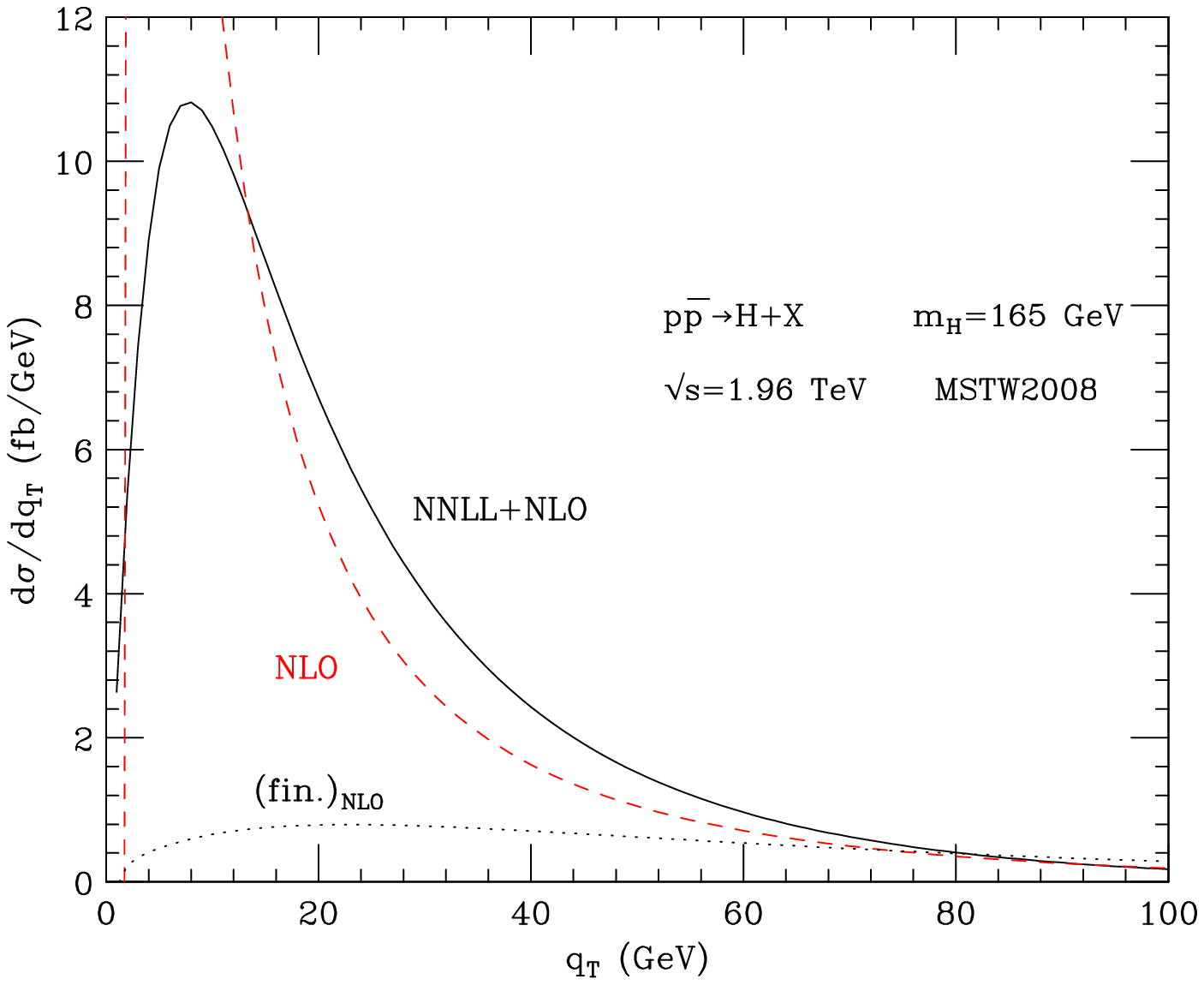}}
\subfigure[]{\label{fig1c}
\includegraphics[width=0.47\textwidth]{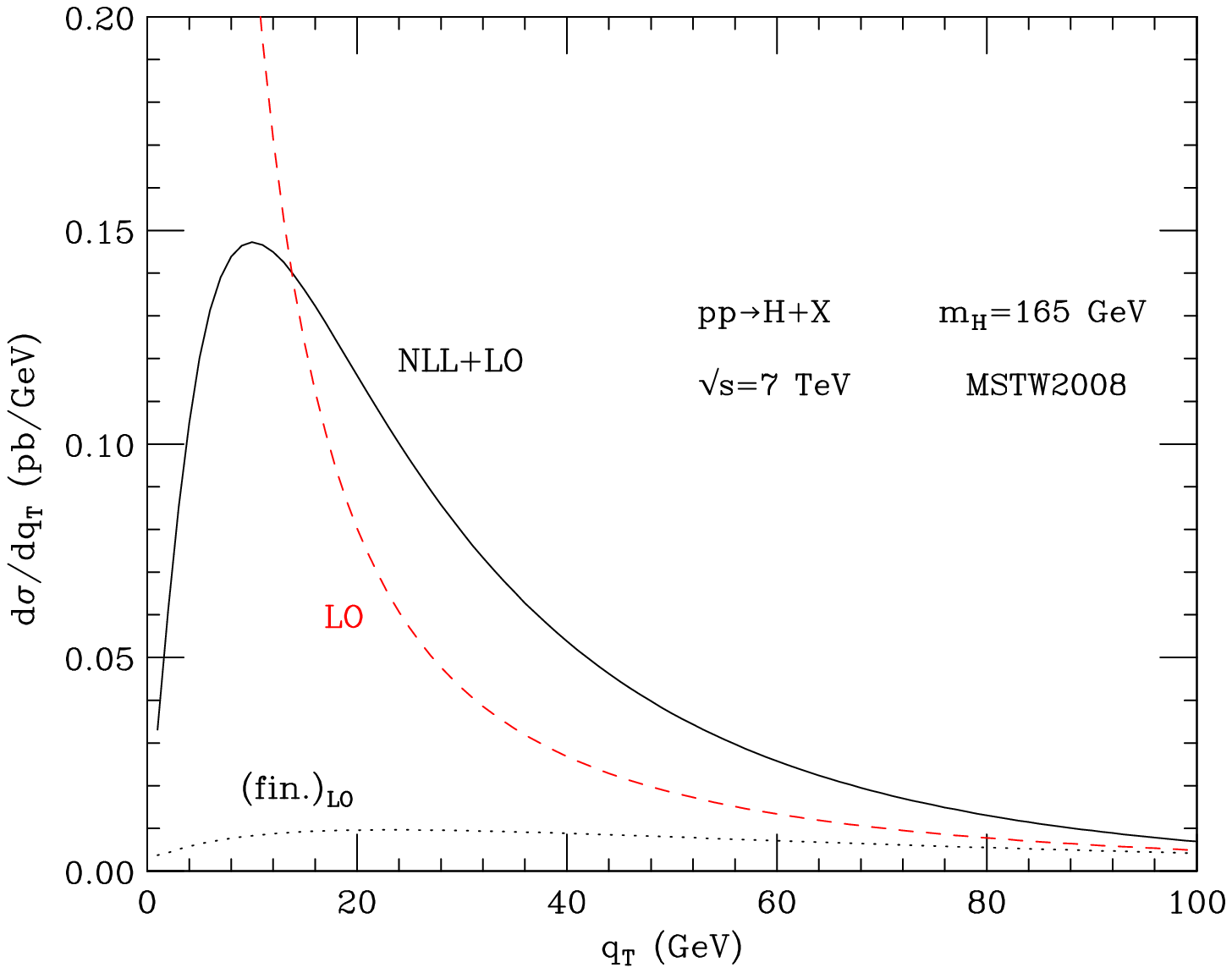}}
\subfigure[]{\label{fig1d}
\includegraphics[width=0.47\textwidth]{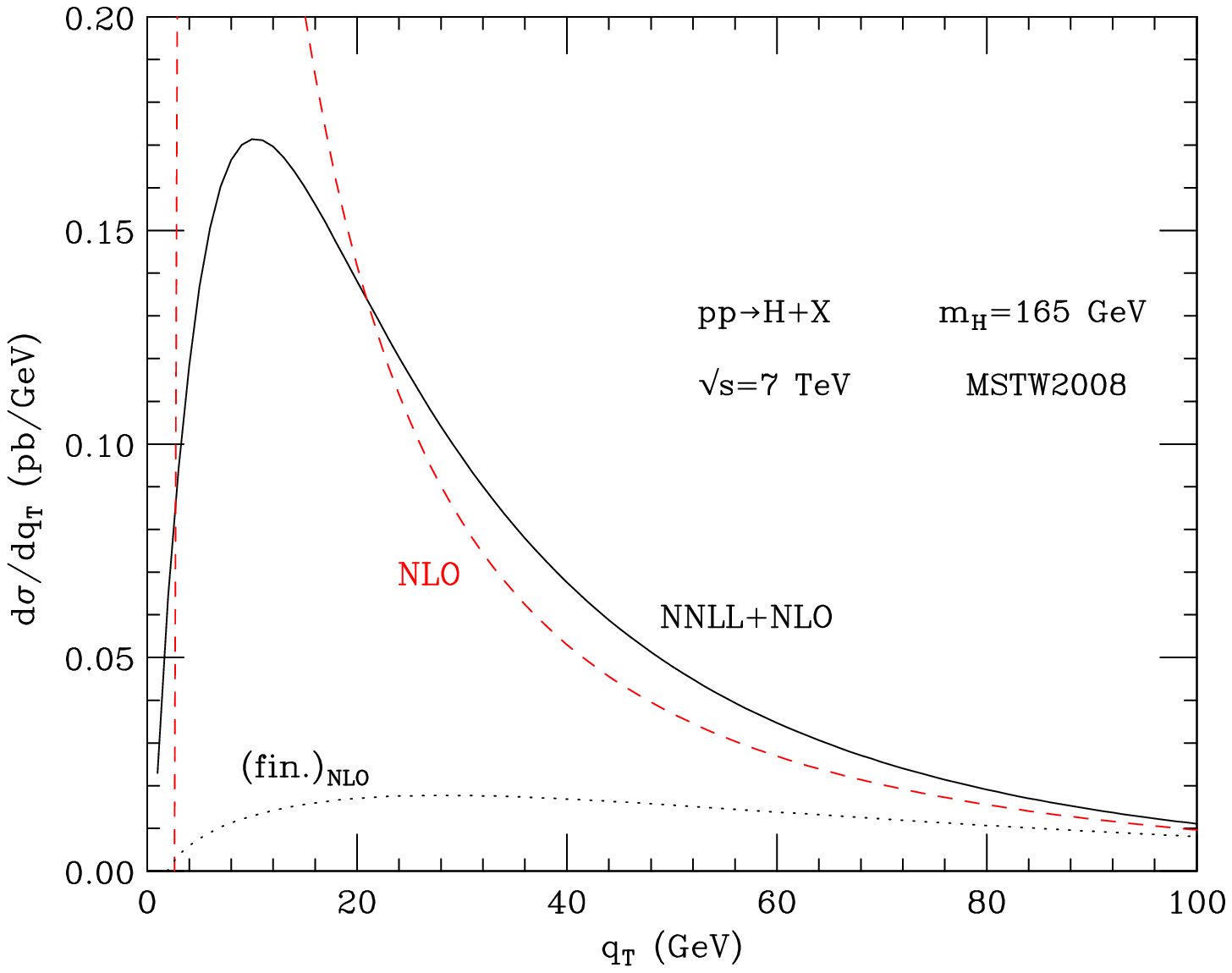}}
\subfigure[]{\label{fig1e}
\includegraphics[width=0.47\textwidth]{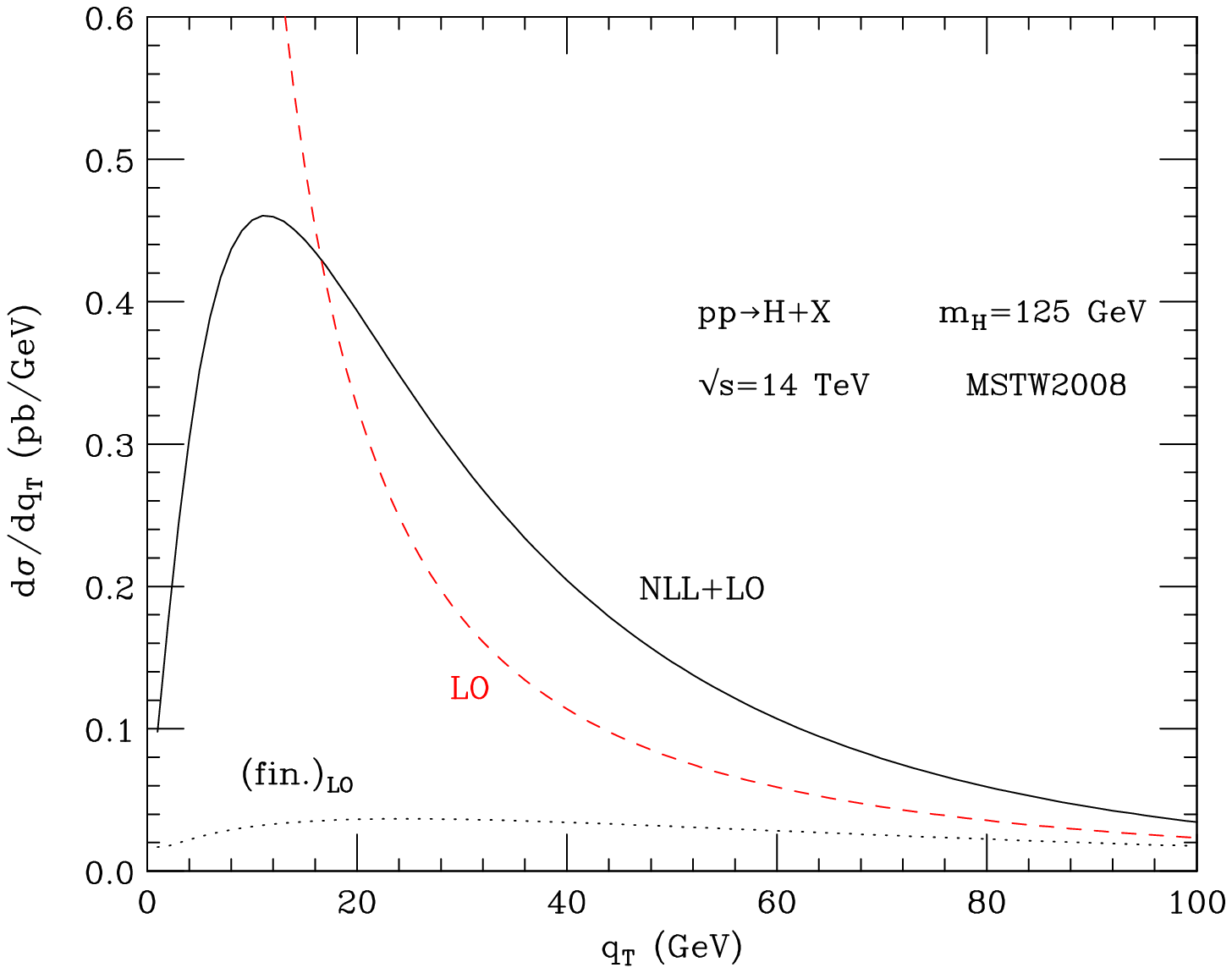}}
\subfigure[]{\label{fig1f}
\includegraphics[width=0.47\textwidth]{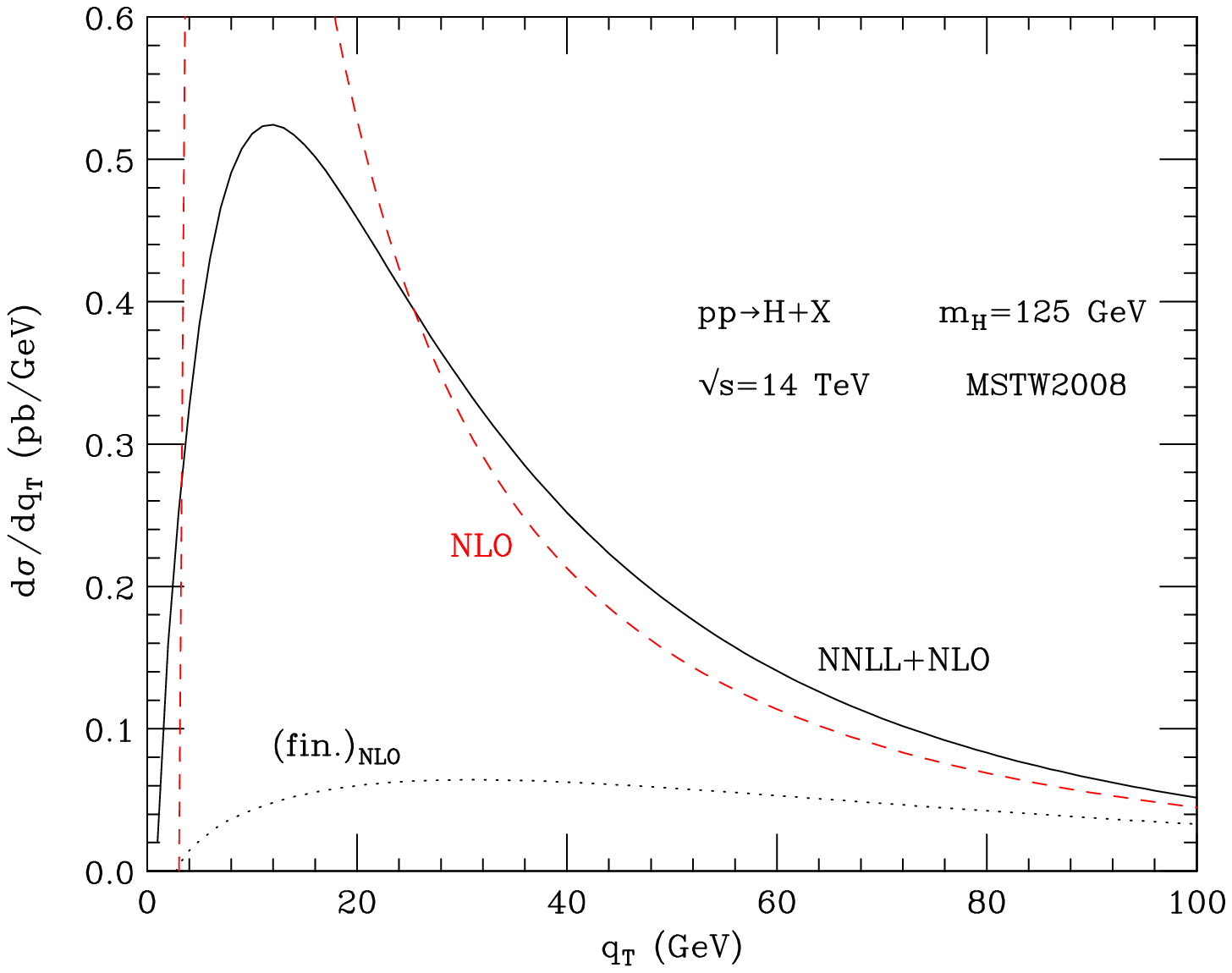}}
\vspace*{-.5cm}
\end{center}
\caption{\label{fig1}
{\em The $q_T$ spectrum of Higgs bosons at the Tevatron and the LHC. Results shown are at NLL+LO (left panels) and NNLL+NLO (right panels)
accuracy. Each result is compared to the corresponding fixed-order result
(dashed line) and to the finite component (dotted line) in Eq.~(\ref{resfin}).}}
\end{figure}


In Fig.~\ref{fig1} (left panels) we present 
the NLL+LO $q_T$ spectrum of a Higgs boson at the Tevatron,
and at the LHC with $\sqrt{s}=7$ TeV and $\sqrt{s}=14$ TeV.
The NLL+LO 
result (solid lines) at
the default scales ($\mu_F=\mu_R=m_H$, $Q=m_H/2$) are compared with the
corresponding
LO results (dashed lines).
The LO finite component of the spectrum (see Eq.~(\ref{resplusfin})) 
is also shown 
for comparison (dotted lines).
We see that the LO result diverges to $+\infty$ as $q_T\to 0$.
The resummation of the small-$q_T$ logarithms
leads to a well-behaved distribution: it vanishes as $q_T\to 0$, has
a kinematical peak,
and tends to the corresponding LO result
at large values of $q_T$.
The finite component
smoothly vanishes
as $q_T\to 0$ but gives a sizable contribution to the NLL+LO result in the low-$q_T$ region.

The results in the right panels of Fig.~\ref{fig1}
are analogous to those in the left panels
although systematically at one order
higher. The $q_T$ spectrum at NNLL+NLO accuracy (solid line) is compared with
the NLO result (dashed line) and with the NLO finite component of the spectrum
(dotted line).
The NLO result diverges to $-\infty$ as $q_T\to 0$ and, at small values of $q_T$,
it has an unphysical peak (the top of the peak is above the vertical scale of the plot)
that is produced
by the numerical compensation of negative leading
and positive subleading logarithmic contributions.
In the region of intermediate values of $q_T$ (say, around $50$~GeV),
the difference between the NNLL+NLO and NLO results gives a sizable contribution 
with respect to the  NLO finite component. This difference is produced by the logarithmic
terms (at NNLO and beyond NNLO) that are included in the resummed calculation at
NNLL accuracy. At large values of $q_T$ the contribution of
the NLO finite component noticeably 
increases. This behaviour indicates that  
the logarithmic terms are no longer dominant and that the resummed
calculation cannot improve upon the predictivity of the fixed-order expansion. 

Comparing the left and right panels of Fig.~\ref{fig1}, we see that
the size of the $q_T$ spectrum increases at NNLL+NLO accuracy with respect to
the  NLL+LO accuracy.
The height of the peak  at NNLL+NLO is larger than at NLL+LO.
The NNLO total cross section,
which fixes the value of the $q_T$ integral of our NNLL+NLO result,
is larger than the NLO total cross section (by about $30\%$ at the Tevatron and $25\%$ at the LHC).
This is due to the positive contribution of both the NNLO terms at small
$q_T$ (the  ${\cal H}_{N}^{H,(2)}$  coefficient of the the 
 ${\cal H}_{N}^{H}$ function and the $g_N^{(3)}$ function in the Sudakov
form factor) and the NLO finite component at intermediate and large values
of $q_T$.

Comparing Fig.~\ref{fig1a},\ref{fig1b} with
Fig.~\ref{fig1c}, \ref{fig1d} and Fig~\ref{fig1e}, \ref{fig1f} we see that
the spectrum is harder at the LHC than at the Tevatron.
The peak of the NNLL+NLO curve moves from $q_T\sim 8$ GeV at the
Tevatron, to $q_T\sim 10$ GeV at the LHC at $\sqrt{s}=7$ TeV, to $q_T\sim 12$ GeV at the LHC at $\sqrt{s}=14$ TeV.


\begin{figure}[htp]
\begin{center}
\subfigure[]{\label{fig2a}
\includegraphics[width=0.48\textwidth]{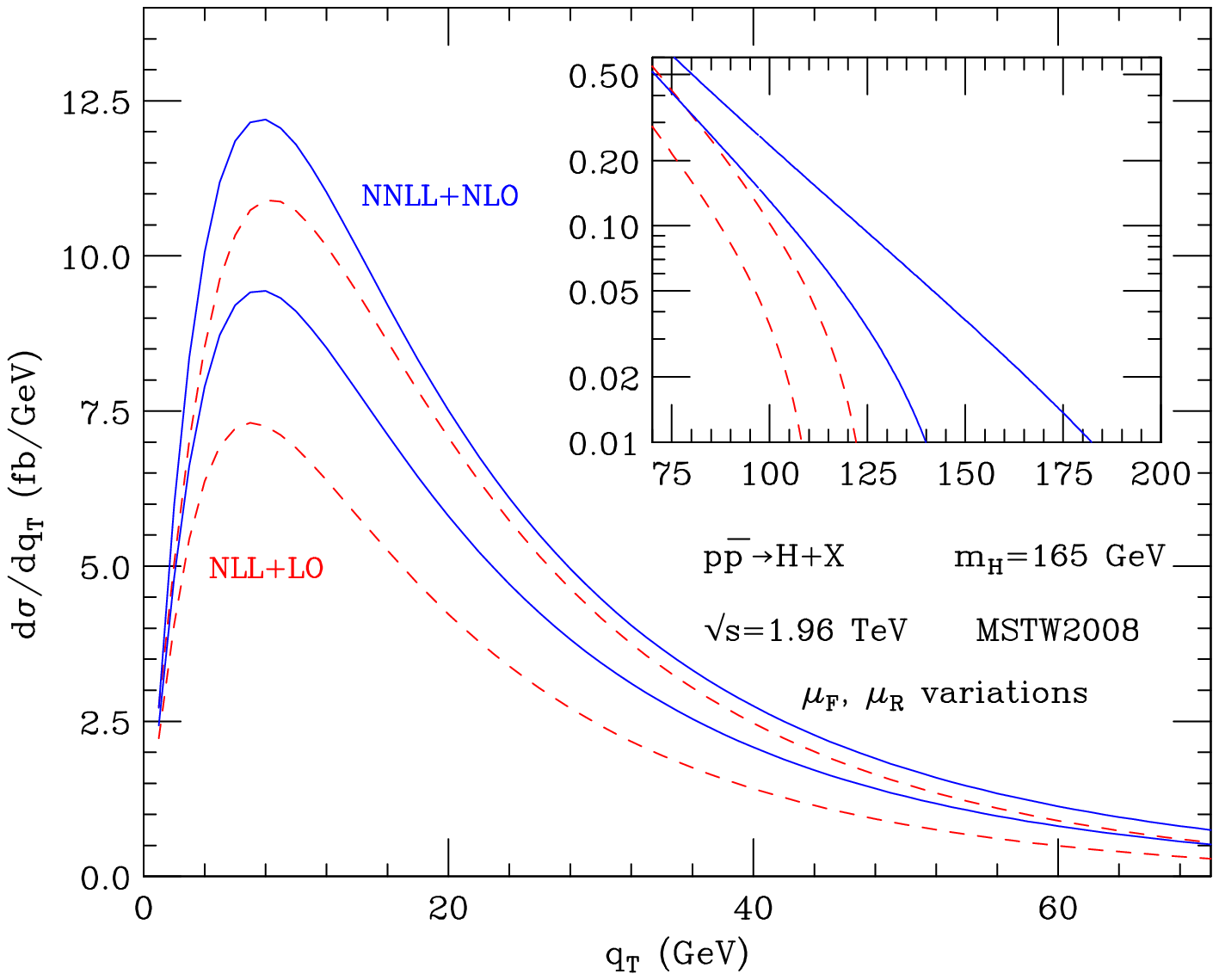}}
\subfigure[]{\label{fig2b}
 \includegraphics[width=0.48\textwidth]{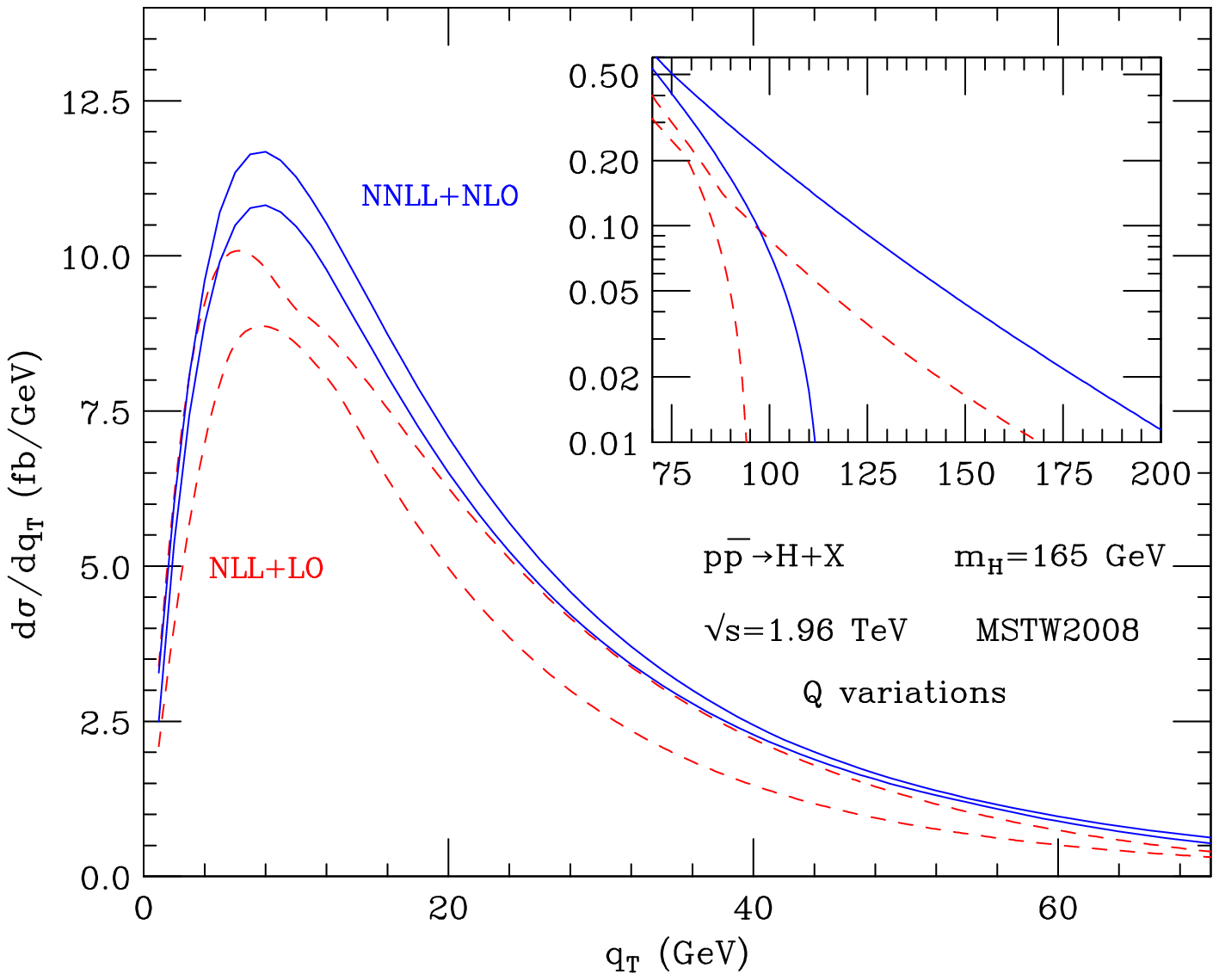}}
\subfigure[]{\label{fig2c}
\includegraphics[width=0.48\textwidth]{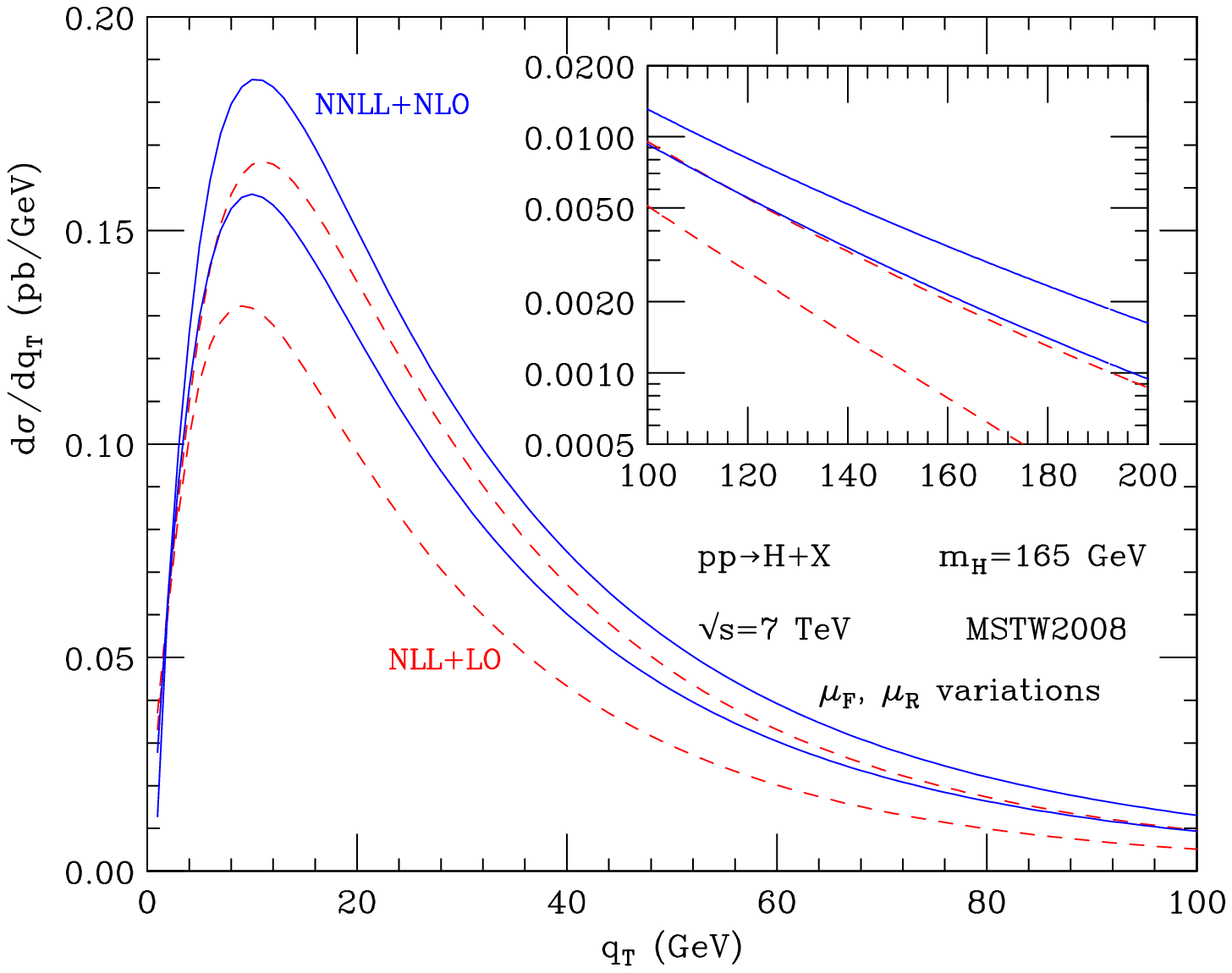}}
\subfigure[]{\label{fig2d}
 \includegraphics[width=0.48\textwidth]{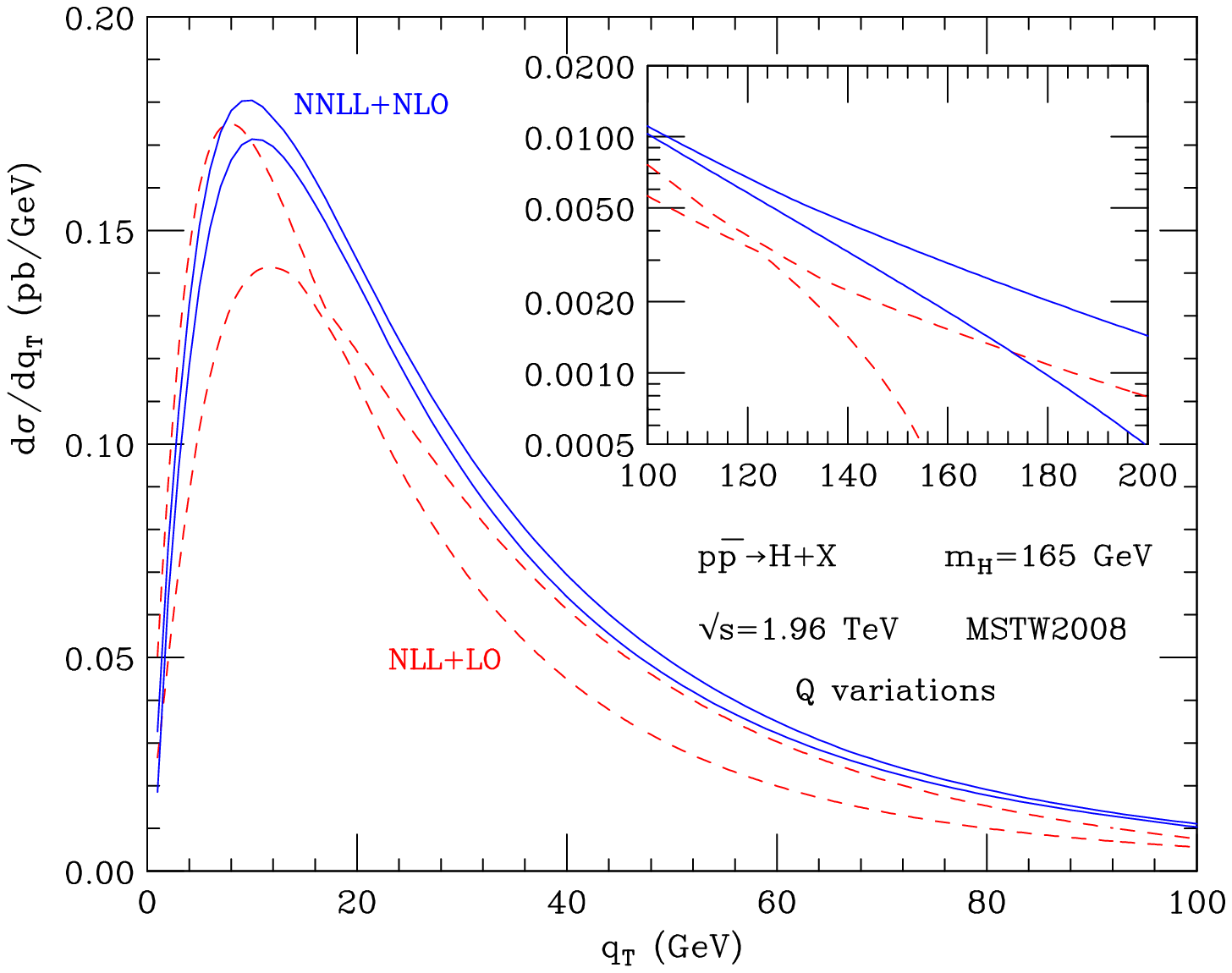}}
\subfigure[]{\label{fig2e}
\includegraphics[width=0.48\textwidth]{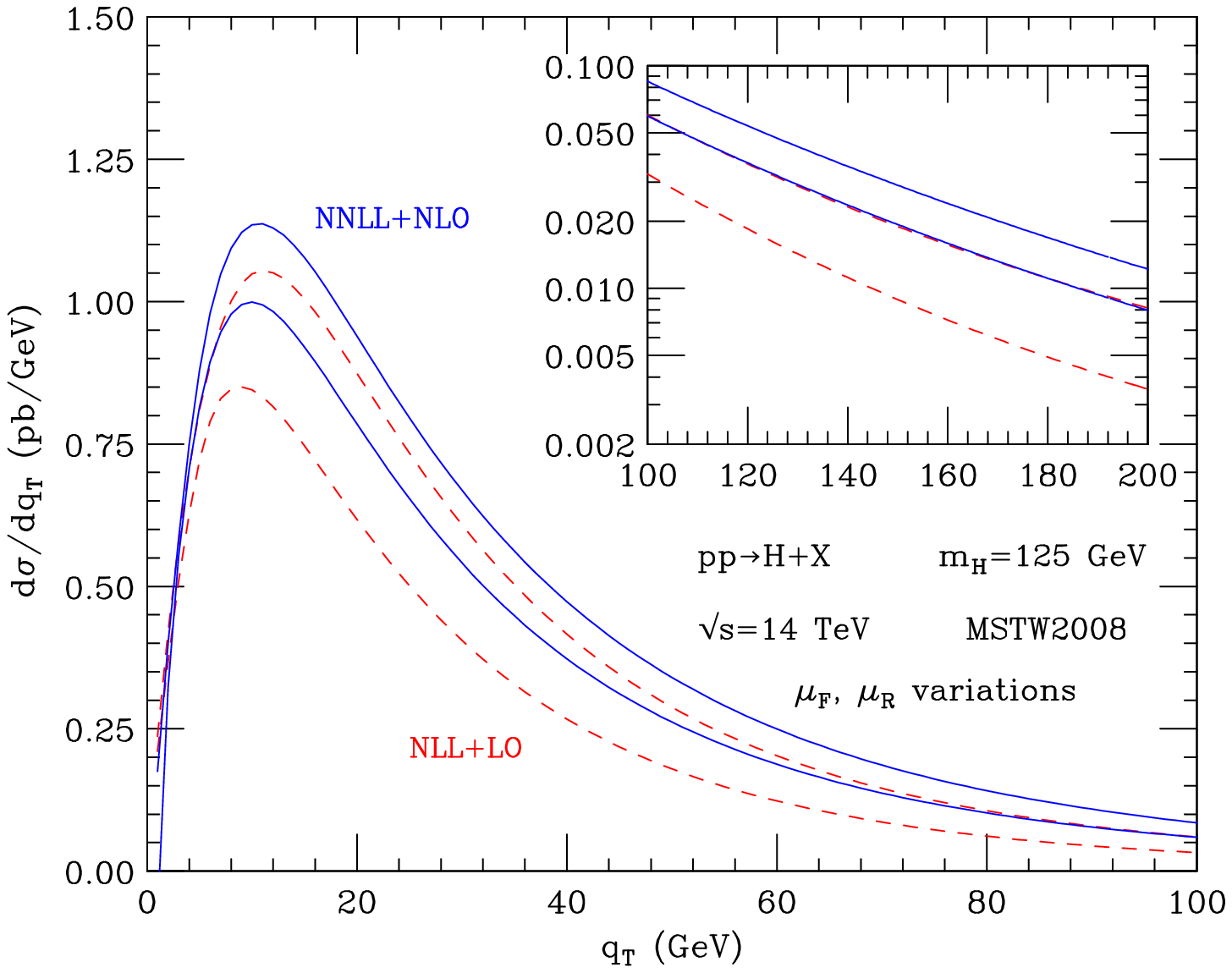}}
\subfigure[]{\label{fig2f}
\includegraphics[width=0.48\textwidth]{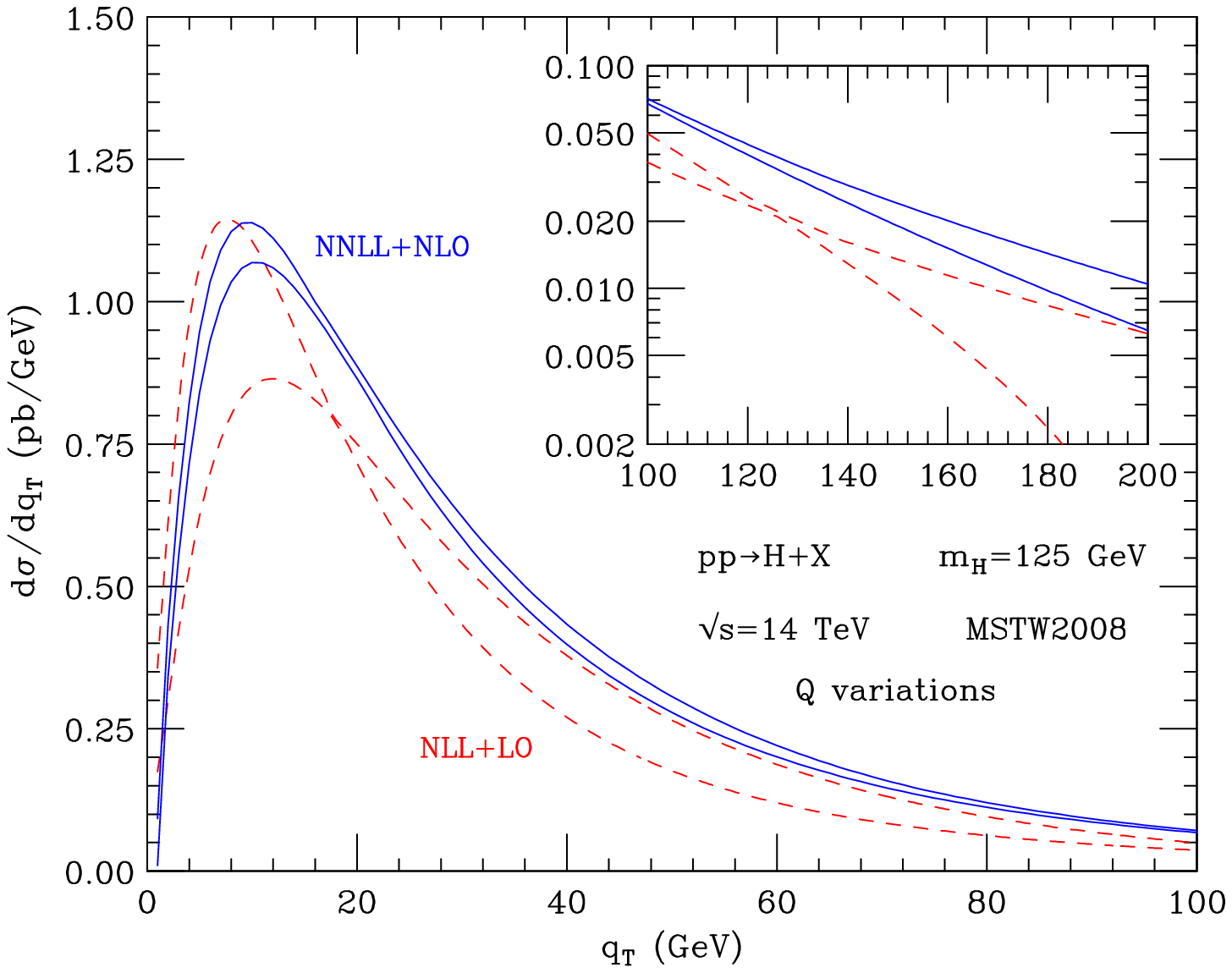}}
\vspace*{-.5cm}
\end{center}
\caption{\label{fig2}
{\em The $q_T$ spectrum of Higgs bosons at the Tevatron and the LHC. The bands are obtained by varying $\mu_F$ and $\mu_R$ (left panels) and $Q$ (right panels) as described in the text.}}
\end{figure}
 
In Fig.~\ref{fig2} we 
show the scale dependence
of the NLL+LO (dashed lines) and NNLL+NLO (solid lines) results.
In the left panels we consider variations of the renormalization and factorization scales. 
The bands are obtained by varying $\mu_R$ and $\mu_F$ as 
previously described in this section.
We note that, in the region of small and intermediate transverse momenta
($q_T\ltap 70$ GeV), the NNLL+NLO and NLL+LO bands overlap. This feature, which is not present in the case of the fixed-order
perturbative results at LO and NLO, 
confirms the importance of 
resummation to achieve a stable perturbative prediction.
In the region of small and intermediate values of $q_T$,
we observe a sensible reduction
of the scale dependence 
going from NLL+LO to NNLL+NLO accuracy.
At the peak the reduction is from  $\pm 20$\% to $\pm 13$\% at the Tevatron,
and from $\pm 11$\% to $\pm 8$\% ($\pm 12$\% to $\pm 7$\%) at the LHC with $\sqrt{s}=7$ ($\sqrt{s}=14$) TeV.
Although $\mu_R$ and $\mu_F$ are varied independently, we find that the 
dependence on $\mu_R$ dominates at any value of $q_T$.

We point out that the $q_T$ region where resummed perturbative predictions are
definitely significant is a wide region from intermediate to
relatively-small (say, close to the peak of the distribution) values of 
$q_T$.
In fact, at very
small values of $q_T$ (e.g. $q_T\ltap 10$~GeV) the size of 
non-perturbative effects is
expected to be important\footnote{See the discussion at the end of this Section.},
while in the
high-$q_T$ region (e.g. $q_T\gtap m_H$~GeV) the resummation of the 
logarithmic terms
cannot improve the predictivity of the fixed-order perturbative expansion.
The inset plots in the figure show the region
from intermediate to large values of $q_T$.
At large $q_T$, the NLL+LO and NNLL+NLO
results deviate from each other,
and the deviation increases as $q_T$ increases.
As previously  stated,
this behaviour is not particularly worrying since,
in the large-$q_T$ region, the resummed results
loose their predictivity and should be replaced by customary  
fixed-order results.

In the right panels of Fig.~\ref{fig2}
we consider resummation scale variations. The bands are obtained
by fixing $\mu_R=\mu_F=m_H$ and varying $Q$ between $m_H/4$ and $m_H$.
Performing variations of the resummation scale,
we can get further insight on the size of yet uncalculated 
higher-order logarithmic contributions at small and intermediate values of 
$q_T$.
We find that, in the region of the peak, at the Tevatron the 
scale dependence
at NNLL+NLO (NLL+LO)
is about $\pm 4$\% ($\pm 10$\%).
At the LHC with $\sqrt{s}=7$ TeV the scale dependence at NNLL+NLO (NLL+LO)
is about $\pm 3$\% ($\pm 8$\%) and at $\sqrt{s}=14$ it is about
$\pm 3$\% ($\pm 13$\%).

Comparing the left and right panels of Fig.~\ref{fig2},
we see that, in the small and intermediate $q_T$ region,
at NNLL+NLO accuracy,
the factorization and renormalization scale dependence
is definitely larger than 
the resummation scale dependence.

The integral over $q_T$ of the resummed NNLL+NLO (NLL+LO)
spectrum is in agreement (for any values 
of $\mu_R, \mu_F$ and $Q$) with the value of
the corresponding NNLO (NLO) total cross section to better than 1\%,
thus checking
the numerical accuracy of the code. 
We also note that the large-$q_T$ region gives a little contribution to
the total cross section;
therefore, the total cross section
constraint mainly acts as a perturbative constraint on the resummed 
spectrum 
in the region from intermediate
to small values of $q_T$.

\begin{figure}[htp]
\begin{center}
\subfigure[]{\label{fig3a}
\includegraphics[width=0.48\textwidth]{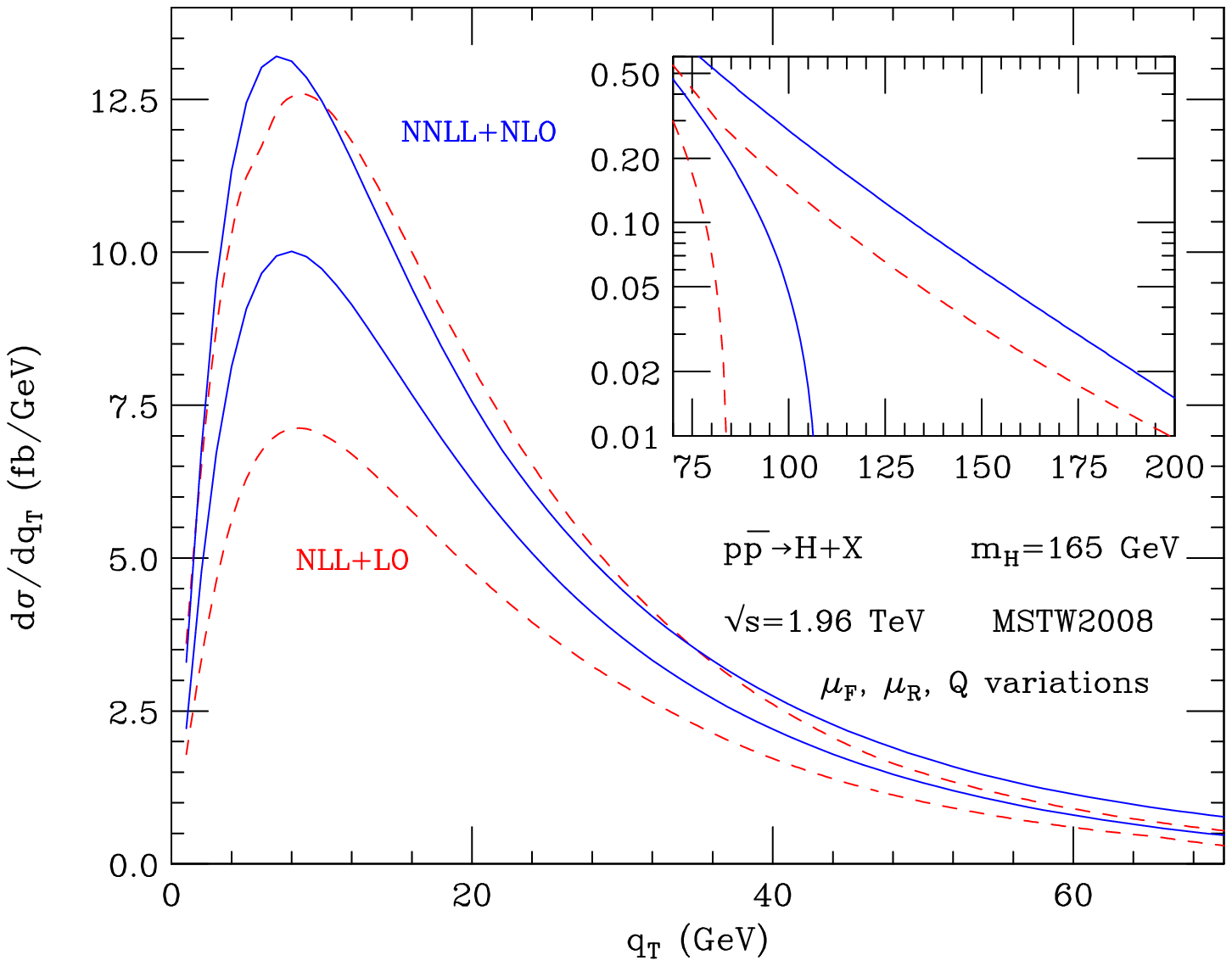} }
\subfigure[]{\label{fig3b}
\includegraphics[width=0.47\textwidth]{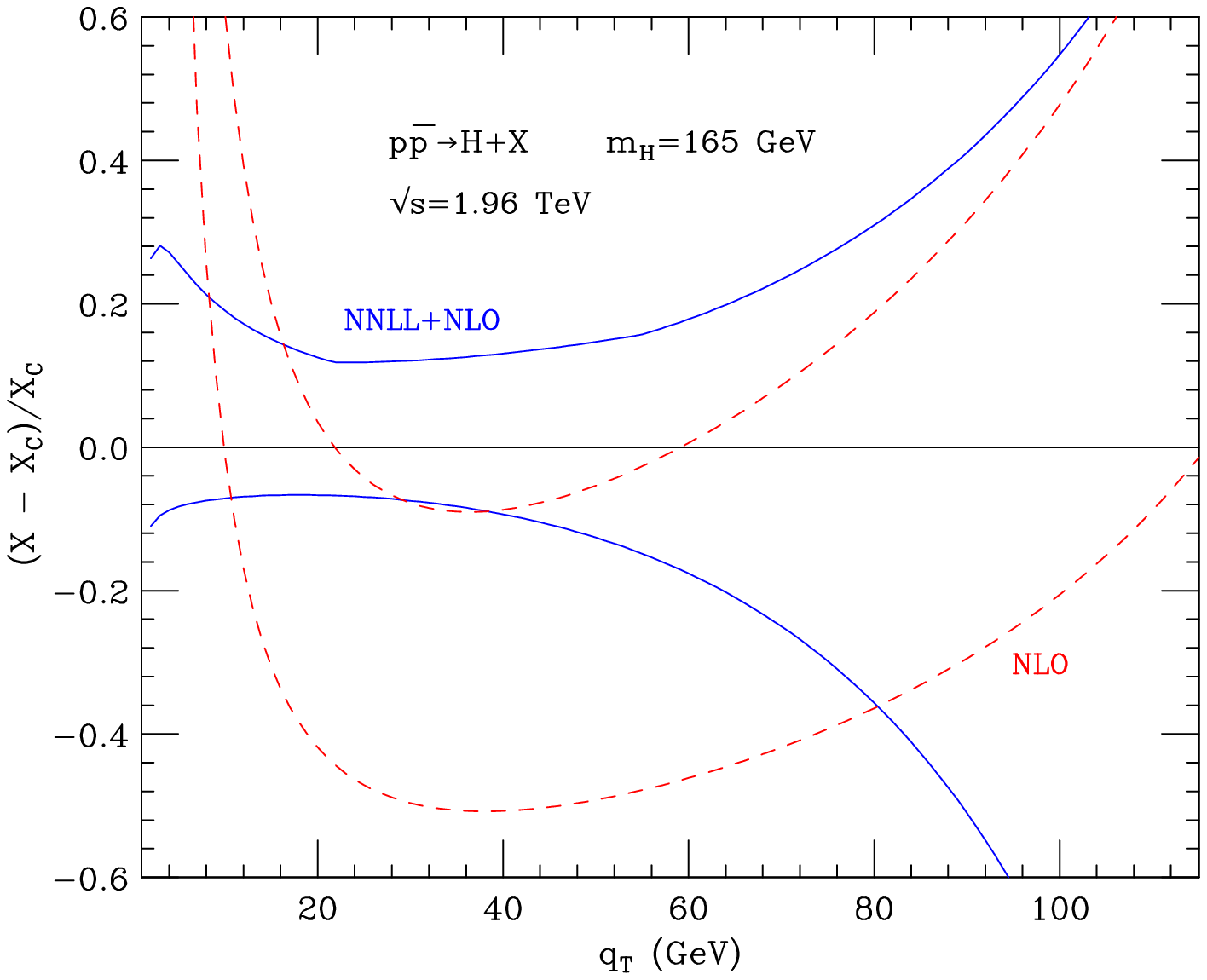}}
\subfigure[]{\label{fig3c}
\includegraphics[width=0.48\textwidth]{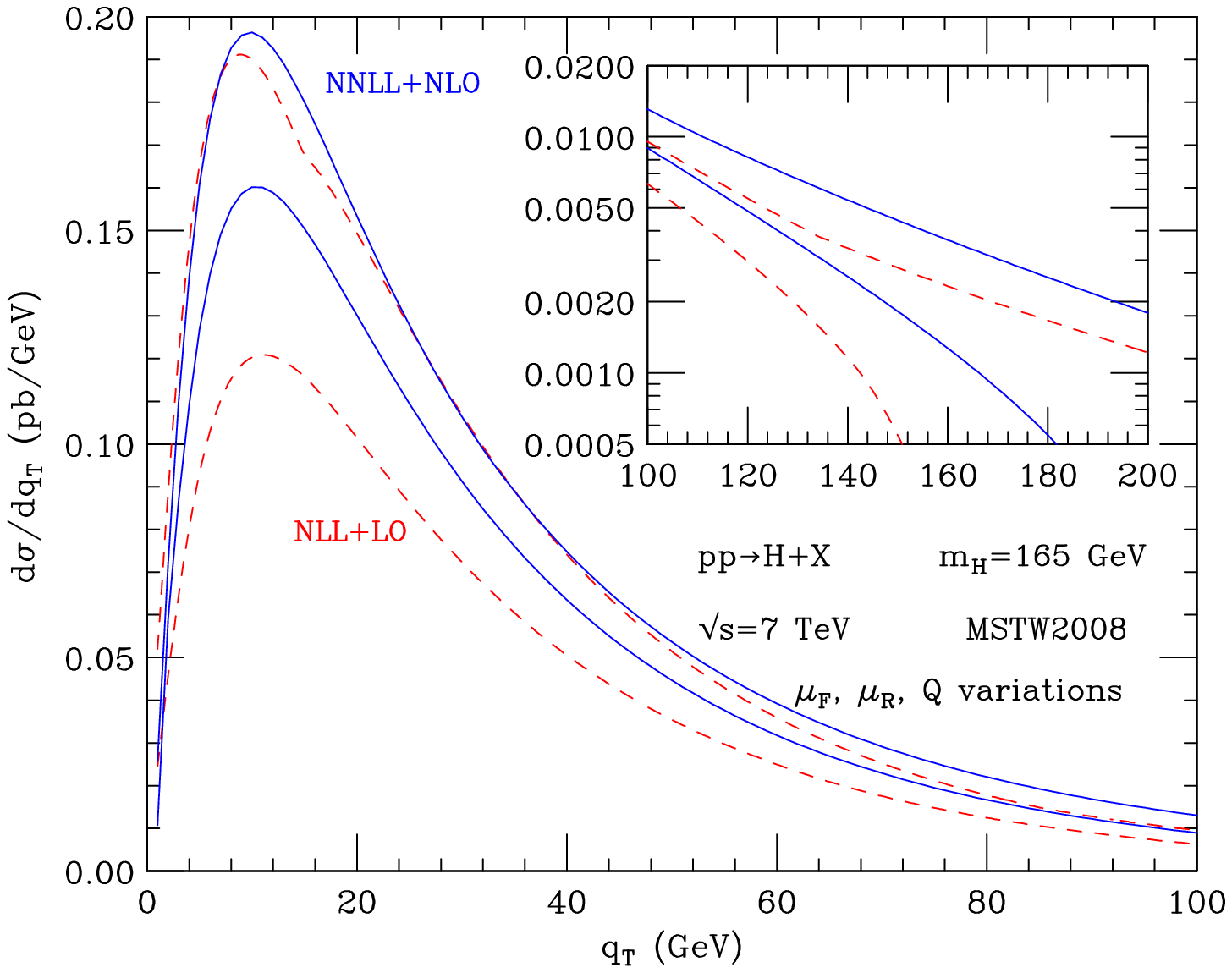} }
\subfigure[]{\label{fig3d}
\includegraphics[width=0.47\textwidth]{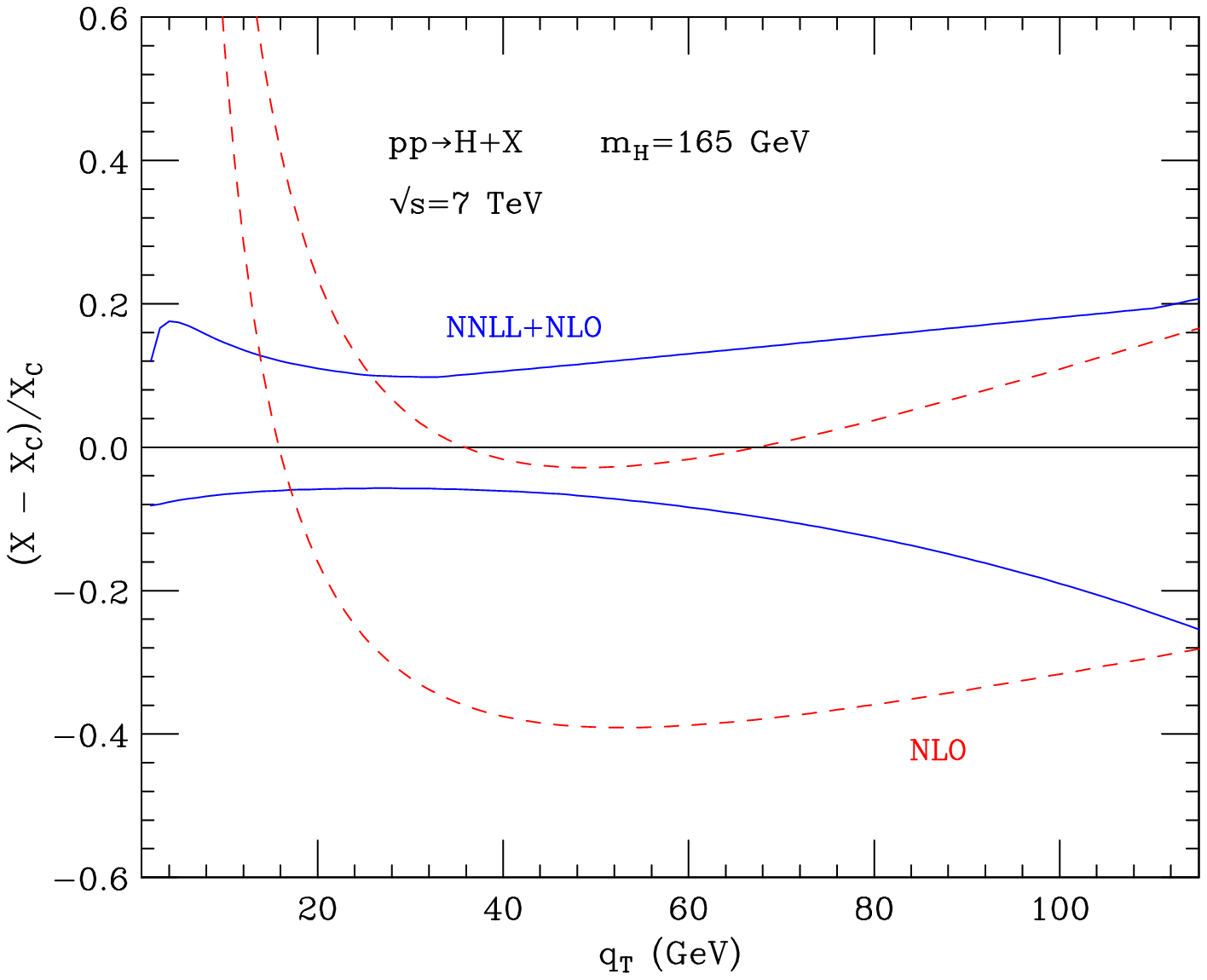}}
\subfigure[]{\label{fig3e}
\includegraphics[width=0.48\textwidth]{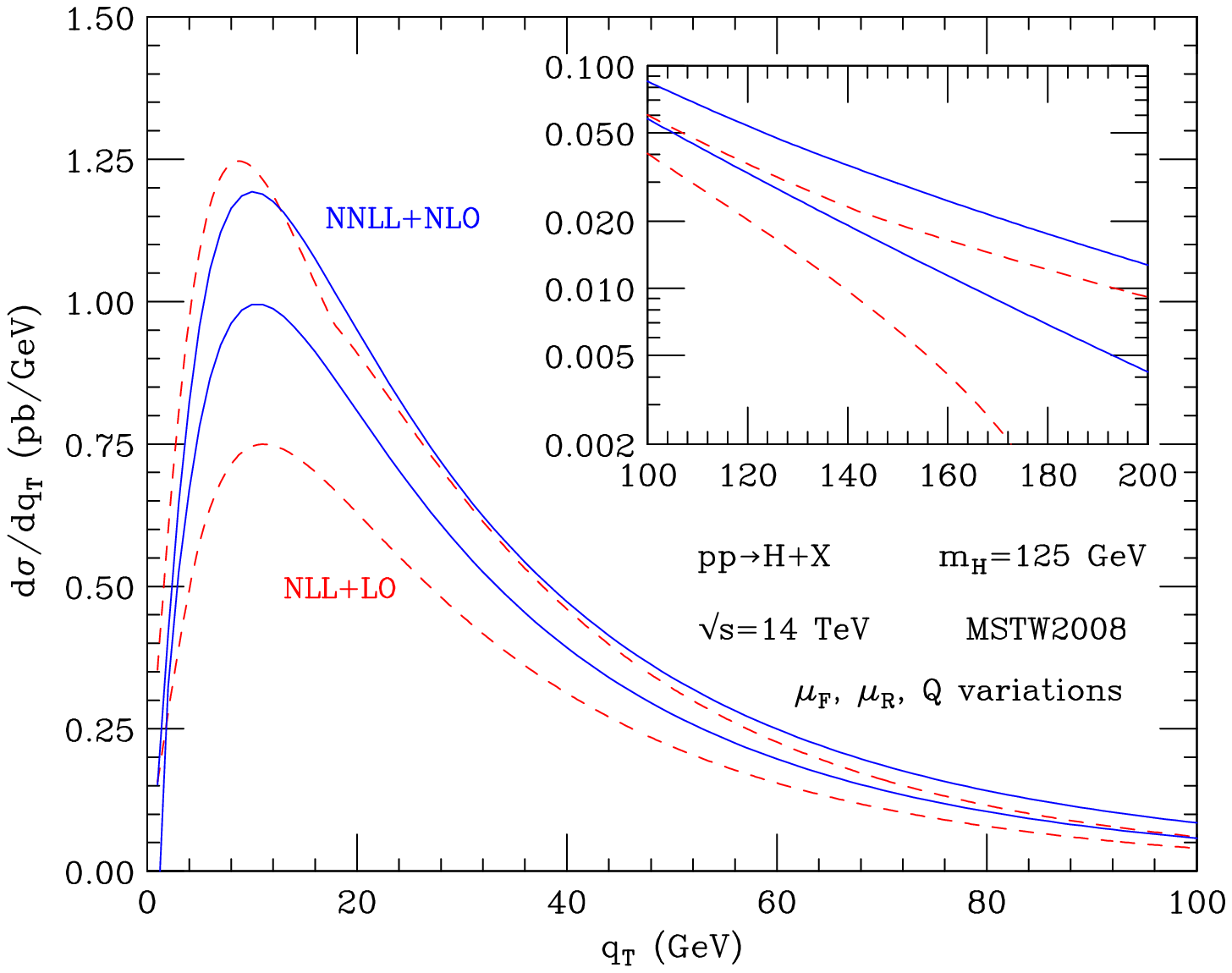}}
\subfigure[]{\label{fig3f}
\includegraphics[width=0.47\textwidth]{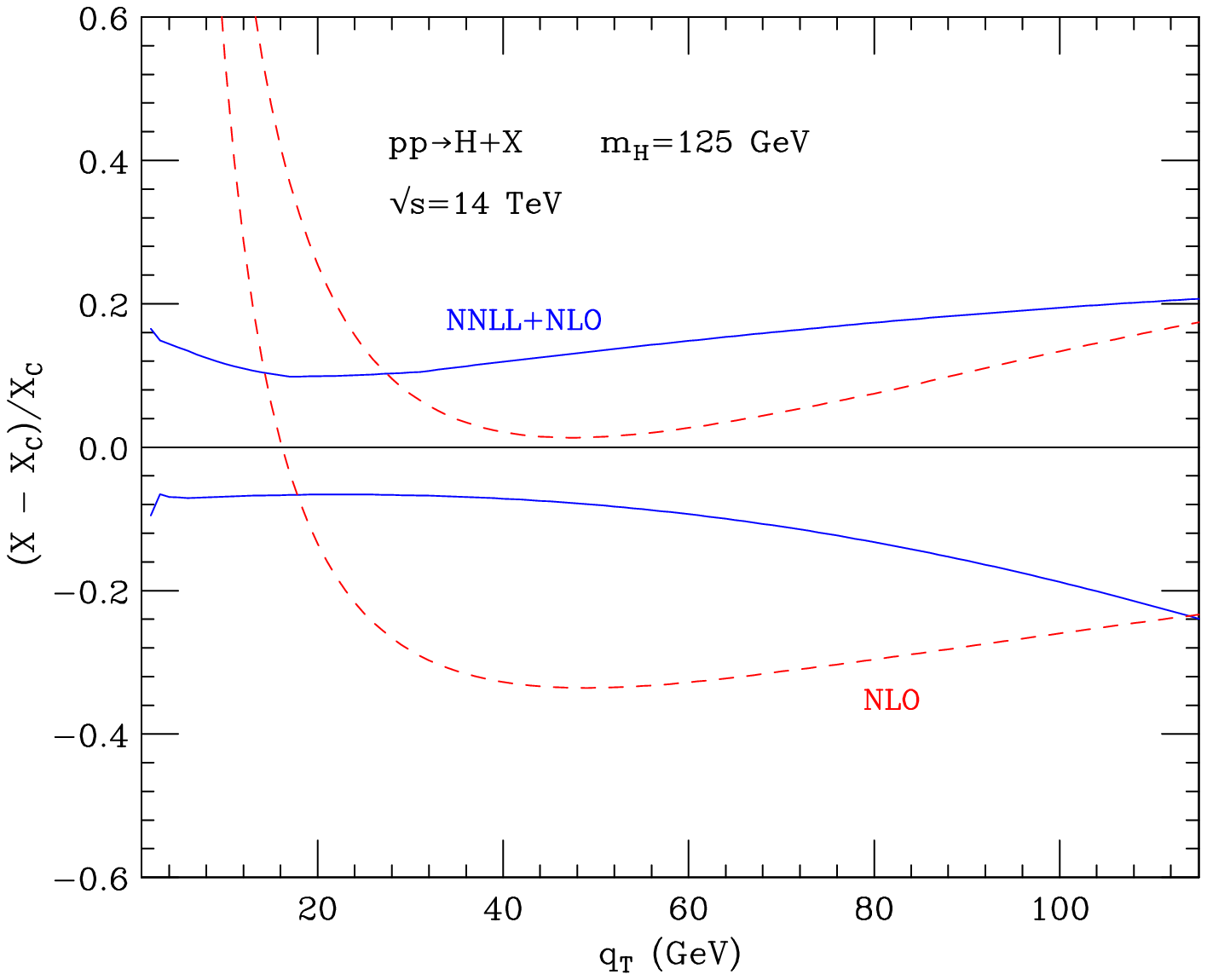}}
\vspace*{-.5cm}
\end{center}
\caption{\label{fig3}
{\em The $q_T$ spectrum of Higgs bosons at the Tevatron and the LHC: 
NNLL+NLO (solid) and NNL+LO (dashes) uncertainty bands (left panels);
NNLL+NLO (solid) and NLO (dashes) uncertainty bands relative to the central NNLL+NLO result (right panels).}}
\end{figure}

In Fig.~\ref{fig3} (left panels)
we report our NLL+LO and NNLL+NLO total scale uncertanty bands
(the inset plots show the large-$q_T$ region).
The bands represent
our best 
estimate of the perturbative uncertainty, and they are obtained by performing
scale variations as follows. We independently vary $\mu_F,\mu_R$ and $Q$
in the ranges
$m_H/2\leq \{\mu_F,\mu_R\} \leq 2m_H$ and $m_H/4\leq Q\leq m_H$,
with the constraints $0.5 \leq \mu_F/\mu_R \leq 2$ and 
$0.5 \leq Q/\mu_R \leq 2$.
The constraint on the ratio $\mu_F/\mu_R$ is the same
as used in Fig.~\ref{fig2}; it has the purpose of avoiding
large logarithmic contributions (powers of $\ln(\mu^2_F/\mu^2_R)$)
that arise from the evolution of the parton densities.
Analogously, the constraint on the ratio $Q/\mu_R$
avoids large logarithmic contributions (powers of $\ln(Q^2/\mu_R^2)$)
in the perturbative expansion of the resummed form factor\,\footnote{We do not apply additional constraints
on the ratio $Q/\mu_F$,
since the form factor
does not depend on $\mu_F$.} $\exp\{{\cal G}_N\}$
(see Eq.~(\ref{exponent})).
We recall (see e.g. Eq.~(19) of Ref. \cite{Bozzi:2005wk})
that
the exponent ${\cal G}_N$ of the form factor is obtained by $q^2$ integration
of perturbative functions of $\as(q^2)$ over the range 
$b_0^2/b^2 \leq q^2\leq Q^2$.
To perform the integration with systematic logarithmic accuracy,
the running coupling $\as(q^2)$ is then expressed in terms of $\as(\mu_R)$ (and $\ln(q^2/\mu_R^2)$).
As a consequence, the renormalization scale $\mu_R$ should not be 
too different from the
resummation scale $Q$, which controls the upper bound of the $q^2$ integration.

A more effective way to show the perturbative uncertainties
is to
consider the fractional difference with respect to a 'reference'
central prediction.
We choose the NNLL+NLO result
at central value of the scales as 
'reference' result, $X_C$, and we show the ratio
$(X-X_C)/X_C$ in  Fig.~\ref{fig3} (right panels). The label $X$
refers to the NNLL+NLO results including scale variations (solid lines),
and to the NLO results including scale variations (dashed lines). 

  
We comment on the overall perturbative uncertainty  band
of our results in Fig.~\ref{fig3} starting from the Tevatron.
The  NNLL +NLO (NLL+LO) uncertainty is about $\pm$13\% 
($\pm$28\%) at the peak, it decreases to about
$\pm$10\% ($\pm$23\%) in the region up to $q_T=30$~GeV, and becomes 
$\pm$18\% ($\pm$20\%) at $q_T=60$~GeV.
In the region beyond $q_T\sim 80$~GeV the resummed
result looses predictivity, and its perturbative uncertainty 
becomes large.

In Fig.~\ref{fig3b} the scale variation band
of the NLO result is compared to the NNLL+NLO band.
The NLO band is obtained by varying $\mu_F$ and $\mu_R$ as for
the NNLL+NLO calculation (the NLO calculation does not depend on the resummation scale $Q$). We see that at large values of $q_T$ the NLO and NNLL+NLO bands
overlap, and the NLO result has smaller uncertainty. As $q_T$ becomes smaller
than about $80$ GeV, the NNLL+NLO has a smaller uncertainty, and the bands
marginally overlap. In this region of transverse momenta, the effect of resummation starts to set in. When $q_T$ becomes smaller and smaller, the NLO band quickly deviates from the NNLL+NLO band and the NLO result becomes unreliable.


We now consider the perturbative uncertainty at the LHC, $\sqrt{s}=7$ TeV.
The  NNLL +NLO (NLL+LO) uncertainty is about $\pm 10$\% 
($\pm 22$\%) at the peak, it decreases to about
$\pm 8$\% ($\pm 19$\%) in the region up to $q_T=30$~GeV, and 
becomes $\pm 10$\% ($\pm 18$\%) at $q_T=60$~GeV. 
In the region beyond $q_T\sim 120$~GeV the resummed
result looses predictivity, and its perturbative uncertainty 
becomes large.
In Fig.~\ref{fig3d} we compare the NLO and NNLL+NLO bands.
The qualitative features are similar to Fig.~\ref{fig3b}: at large values of
$q_T$ the NLO and NNLL+NLO scale uncertainty bands
overlap, and the NLO result has smaller uncertainty. As $q_T$ becomes smaller
than about $120$ GeV, the NNLL+NLO has a smaller uncertainty, but the bands
still overlap. In the region of intermediate transverse momenta ($q_T\sim 50$ GeV), the bands marginally overlap and the NLO result underestimates the
cross section.  When $q_T$ becomes smaller, the NLO band quickly deviates from the NNLL+NLO band and the NLO result becomes unreliable.


We finally consider the perturbative uncertainty at the LHC when $\sqrt{s}=14$ TeV.
The  NNLL +NLO (NLL+LO) uncertainty is about $\pm 9$\% 
($\pm 25$\%) at the peak, it decreases to about
$\pm 8$\% ($\pm 19$\%) in the region up to $q_T=30$~GeV, and 
moves to $\pm 12$\% ($\pm 19$\%) at $q_T=60$~GeV. 
In the region beyond $q_T\sim 150$~GeV the resummed
result looses predictivity, and its perturbative uncertainty 
becomes large.
In Fig.~\ref{fig3f} we compare the NLO and NNLL+NLO scale uncertainty bands.
The qualitative features are similar to those of Figs.~\ref{fig3b}, \ref{fig3d}:
at large values of $q_T$ the NLO and NNLL+NLO bands overlap and the NLO result has
smaller uncertainty. In the region of intermediate transverse momenta ($q_T\sim 50$ GeV), the bands marginally overlap and the NLO result underestimates the
cross section.  When $q_T$ becomes smaller, the NLO result becomes unreliable.

Comparing Fig.~\ref{fig3a},\ref{fig3b} with Fig.~\ref{fig3c},\ref{fig3d}
and Fig.~\ref{fig3e},\ref{fig3f} we see that perturbative uncertainties are larger
at the Tevatron than at the LHC. We also note that our NNLL+NLO result is much more stable at the LHC than at the Tevatron, where its validity is confined to a smaller
region of transverse momenta. This is not completely unexpected.
At smaller values of the center of mass energy, the production of the Higgs boson
is accompanied by softer radiation, and thus the $q_T$ spectrum is softer than at
the LHC.

We conclude this section with a discussion on the uncertainties on the normalized $q_T$ spectrum (i.e., $1/\sigma \times d\sigma/dq_T$). As mentioned in the introduction, the
typical procedure of the experimental collaborations is to use the information
on the total cross section \cite{Dittmaier:2011ti} to rescale the best theoretical predictions of Monte Carlo
event generators, whereas the
NNLL+NLO result of our calculation, obtained with the public program {\ttfamily HqT},
is used to reweight the transverse-momentum
spectrum of the Higgs boson obtained in the simulation.
Such a procedure implies that the important information provided by the resummed NNLL+NLO spectrum
is not its integral, i.e. the total cross section, but its {\em shape}.
The sources of uncertainties on the shape of the spectrum are essentially the same
as for the inclusive cross section: the uncertainty from missing higher-order contributions,
estimated through scale variations, and PDF uncertainties.
One additional uncertainty in the $q_T$ spectrum that needs be considered comes from Non-Perturbative (NP) effects.

We remind the reader that the quantitative predictions presented in this paper are obtained in a purely perturbative framework.
It is known \cite{Dokshitzer:hw} that the transverse-momentum
distribution is affected by NP effects, which become important as $q_T$ becomes small.
A customary way of modelling these effects is to introduce an NP transverse-momentum smearing
of the distribution. In the case of resummed calculations in impact parameter space,
the NP smearing is implemented by multiplying the $b$-space perturbative
form factor by an NP form factor.
The parameters controlling this NP form factor are typically obtained through a comparison to data.
Since there is no evidence for the Higgs boson yet, the procedure to fix the NP form factor is somewhat arbitrary.
Here we follow the procedure adopted in Ref.~\cite{Bozzi:2005wk}, and we multiply the resummed form factor in Eq.~(\ref{resum}) by a gaussian smearing $S_{NP}=\exp\{-g b^2\}$, where the parameter $g$ is taken in the range ($g=1.67-5.64$ GeV$^2$) suggested by the study of Ref.~\cite{Kulesza:2003wi}\footnote{We note that the inclusion of
this smearing factor does not change the overall normalization, since $S_{NP}(b=0)=1$}.
The above procedure can give us some insight on the quantitative impact of these NP effects on the Higgs boson spectrum.

\begin{figure}[htp]
\begin{center}
\subfigure[]{\label{fig4a}
\includegraphics[width=0.46\textwidth]{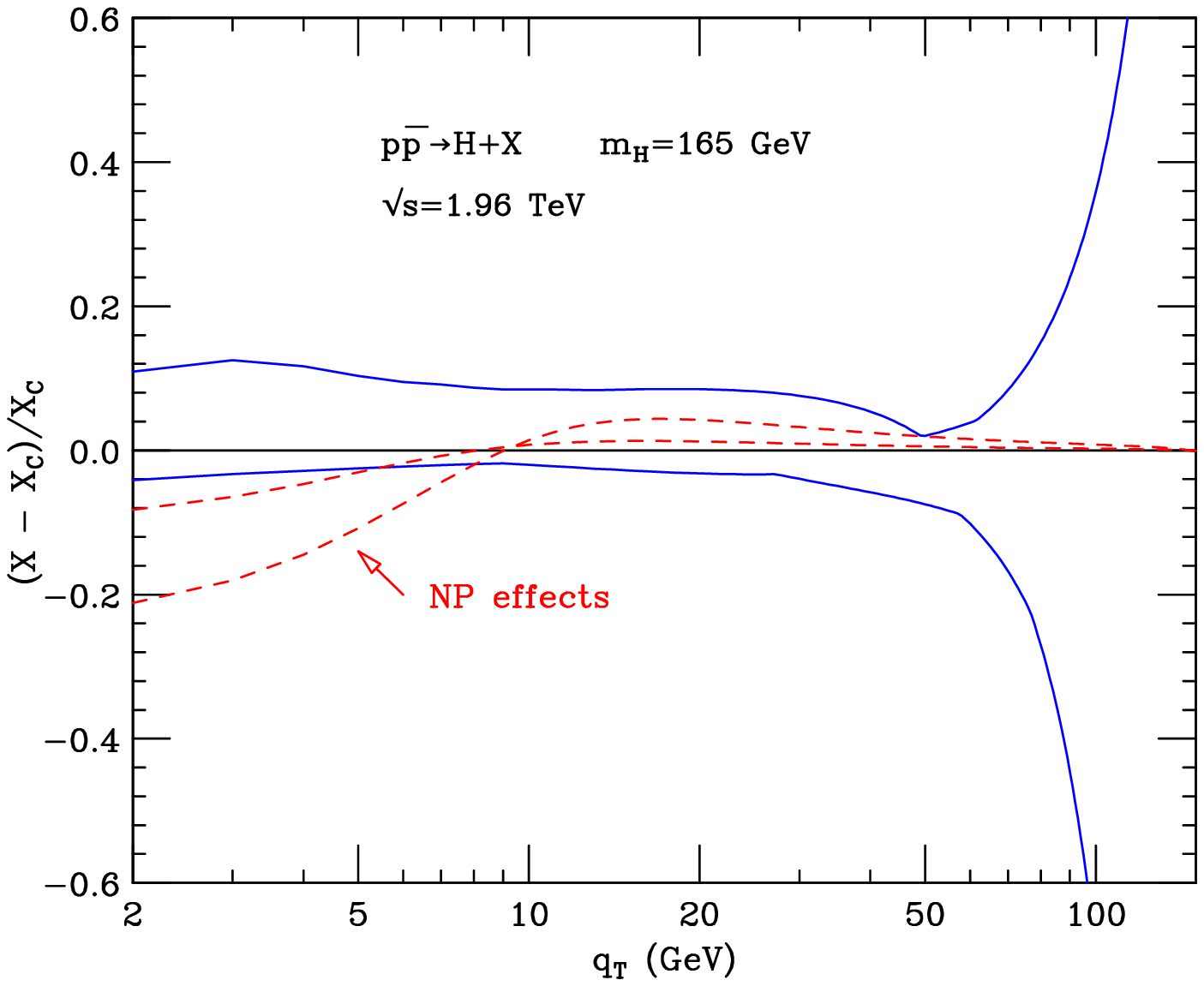} }
\subfigure[]{\label{fig4b}
\includegraphics[width=0.46\textwidth]{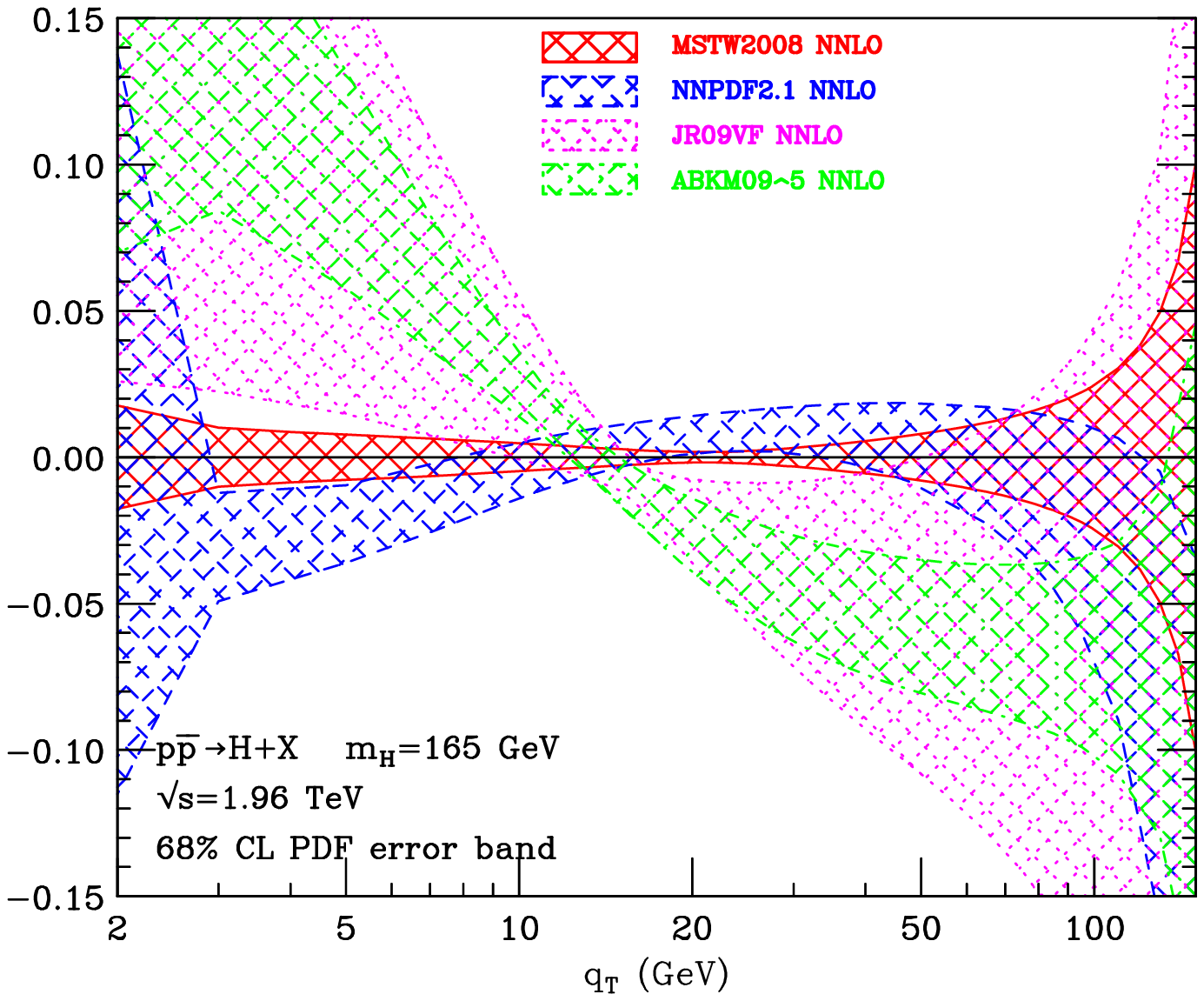}}
\subfigure[]{\label{fig4c}
\includegraphics[width=0.46\textwidth]{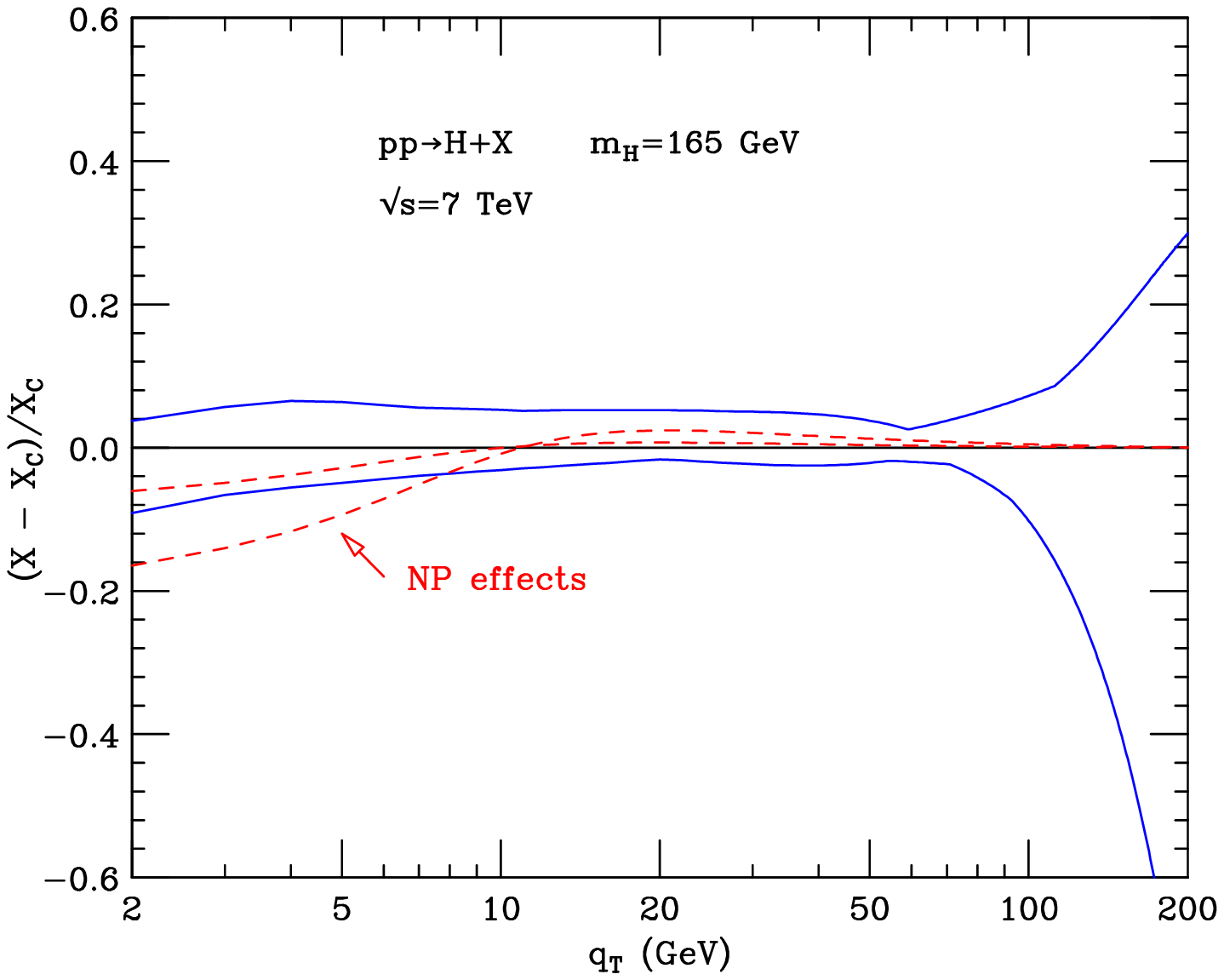} }
\subfigure[]{\label{fig4d}
\includegraphics[width=0.46\textwidth]{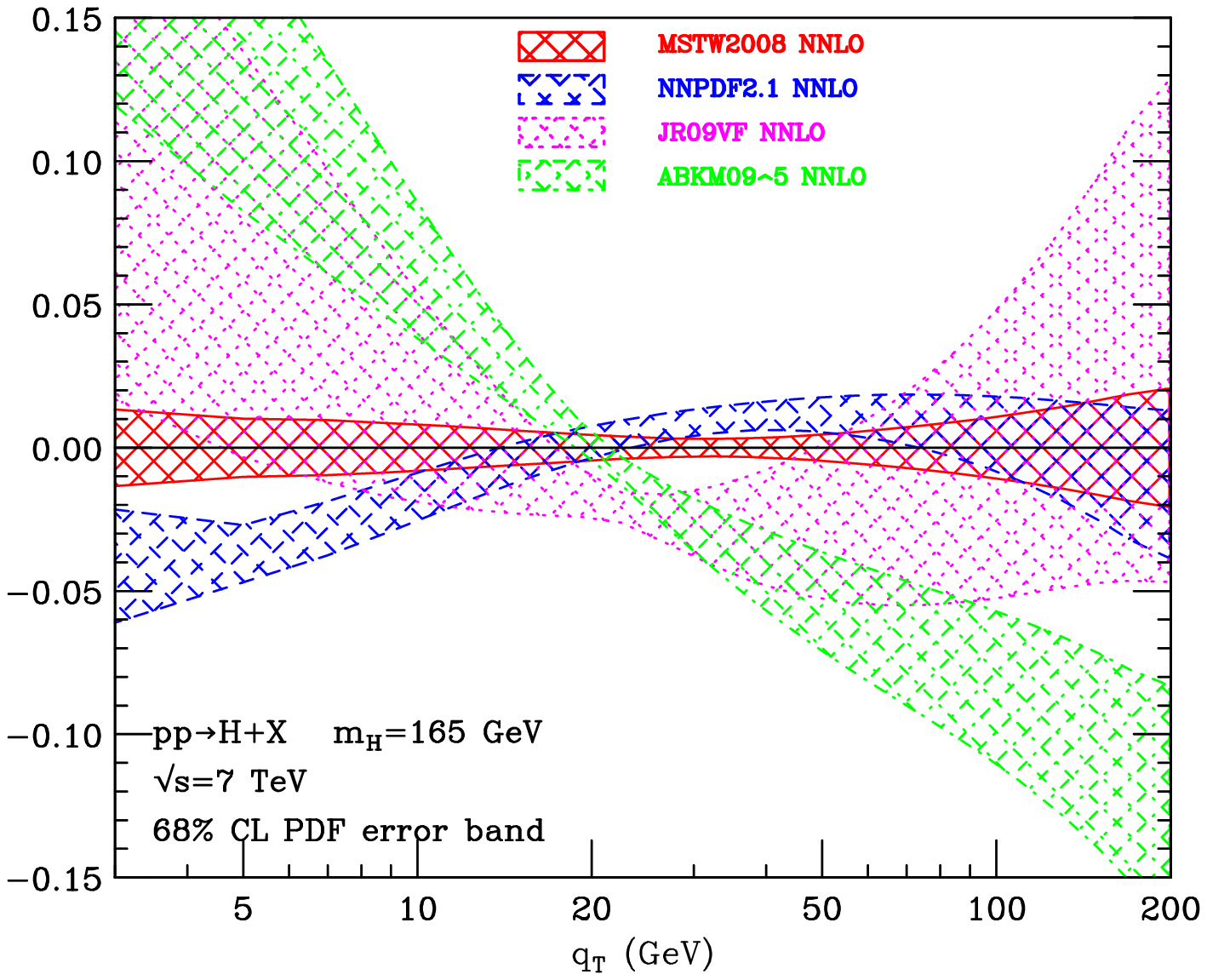}}
\subfigure[]{\label{fig4e}
\includegraphics[width=0.46\textwidth]{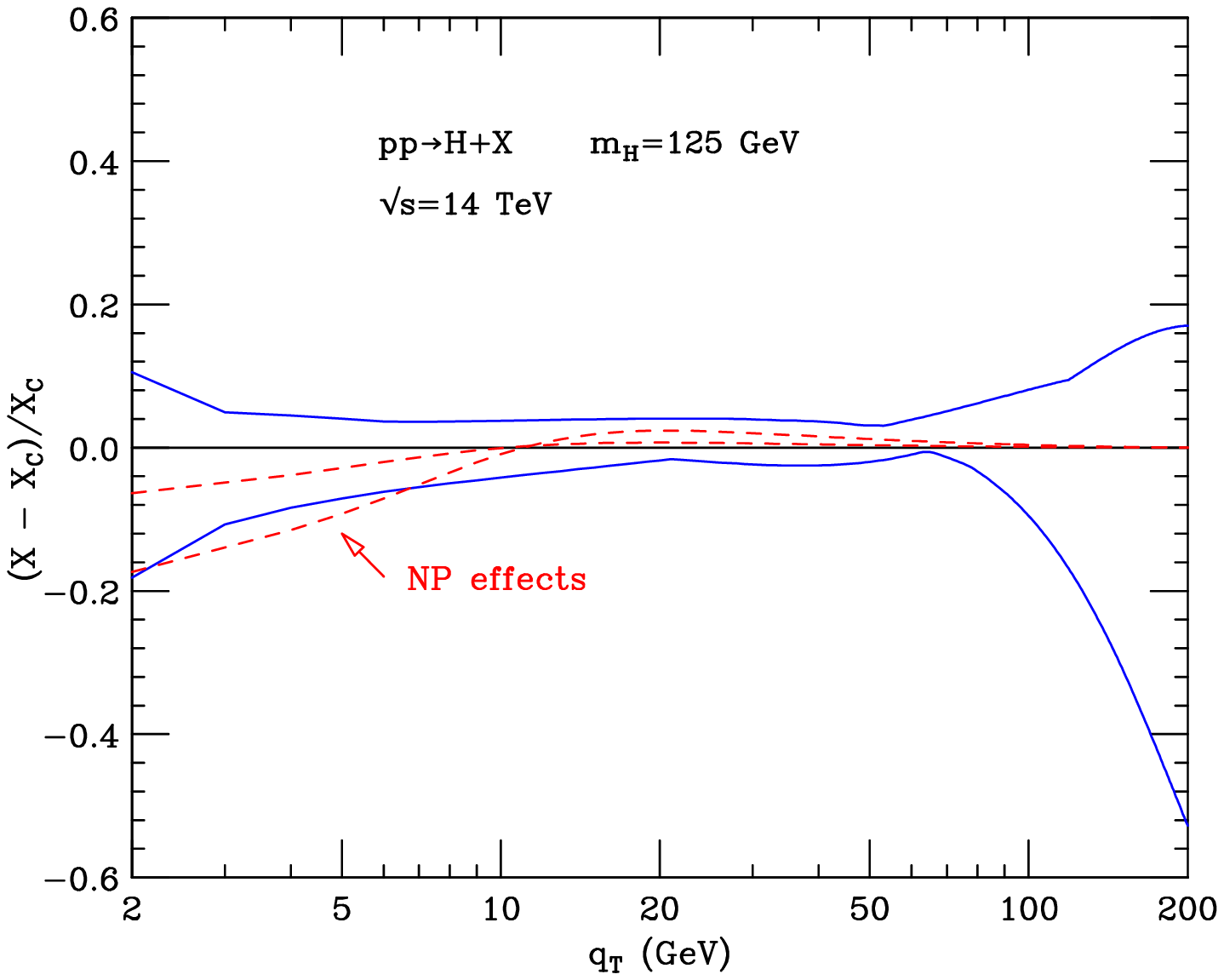}}
\subfigure[]{\label{fig4f}
\includegraphics[width=0.46\textwidth]{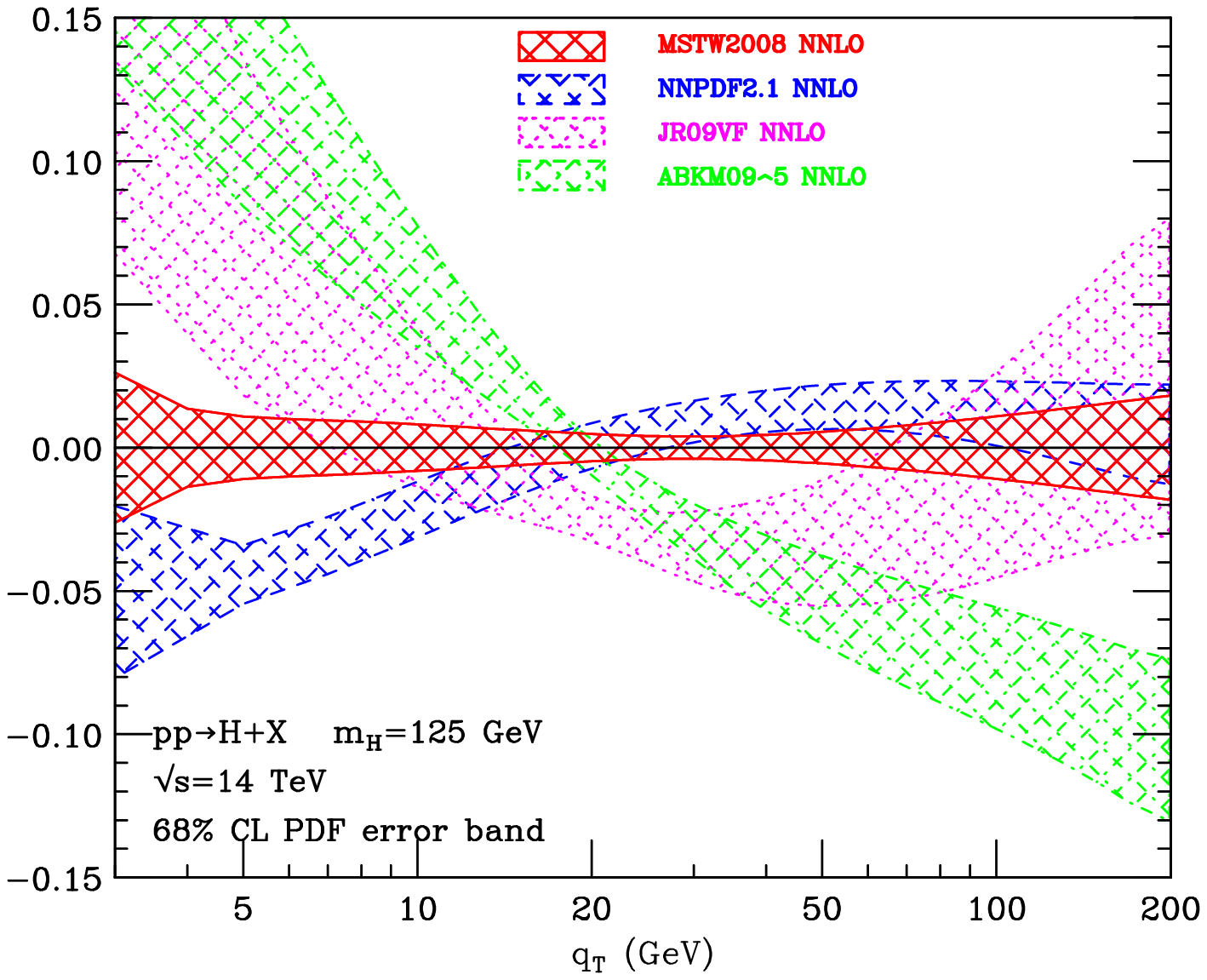}}
\vspace*{-.7cm}
\end{center}
\caption{\label{fig4}
{\em Uncertainties in the normalized $q_T$ spectrum of the Higgs boson at the Tevatron and the LHC. Left panels: the
NNLL+NLO uncertainty bands (solid) computed as in Fig.~3 compared to an estimate of NP effects (dashed).
Right panels: PDF uncertainties bands at 68\% CL. All results are relative to the NNLL+NLO central value computed with MSTW2008 NNLO PDFs.}}
\end{figure}

In Fig.~\ref{fig4} (left panels) we compare the NNLL+NLO shape uncertainty as coming from scale variations (solid lines)
to the NP effects (dashed lines).
The bands are obtained by normalizing each spectrum to unity, and computing the
relative difference with respect to the central normalized prediction obtained
with the MSTW2008 NNLO set (with $g=0$).
A comparison of Fig.~\ref{fig4a},\ref{fig4c},\ref{fig4e} to Fig.~\ref{fig3b},\ref{fig3d},\ref{fig3f} shows that the scale uncertainty on the normalized NNLL+NLO distribution is smaller than the corresponding uncertainty on the NNLL+NLO result. This is not unexpected: a sizeable contribution to the uncertainties
shown in Fig.~\ref{fig3} comes actually from uncertainties on the total cross section, which do not contribute in Fig.~\ref{fig4}.
In other words, studying uncertainties on the normalized distribution allows
us to assess the true uncertainty in the shape of the resummed $q_T$ spectrum.

At the Tevatron (Fig.~\ref{fig4a}) such scale uncertainty ranges from $+8\%-3\%$ in the region of the peak, to $+3\%-8\%$ when $q_T\sim 50$ GeV. At larger values of $q_T$ the uncertainty of the NNLL+NLO resummed distribution increases consistently with the behaviour observed in Fig.~\ref{fig3b}.
The inclusion of the NP effects makes the distribution harder, the effect ranging from 10\% to 20\%
in the very small-$q_T$ region. For $q_T\gtap 10$ GeV the impact of NP effects is of the order of about $5\%$ and decreases as $q_T$ increases. 
At the LHC, $\sqrt{s}=7$ TeV (Fig.~\ref{fig4c}) the scale uncertainty ranges from $+5\%-3\%$ in the region of the peak to $+5\%-4\%$ at $q_T\sim 80$ GeV.
At the LHC, $\sqrt{s}=14$ TeV (Fig.~\ref{fig4e}) the shape uncertainty ranges from $+5\%-3\%$ in the region of the peak to $+8\%-9\%$ at $q_T\sim 100$ GeV.
The impact of NP effects is similar at $\sqrt{s}=7$ and $14$ TeV: it ranges from about 10\% to 20\% in the region below the peak, is about $3-4\%$ for $q_T\sim 20$ GeV, and quickly decreases as $q_T$ increases.
We conclude that the uncertainty from unknown NP effects is smaller than the scale uncertainty,
and is comparable to the latter only in the very small $q_T$ region.

The impact of PDF uncertainties at $68\%$ CL on the shape of the $q_T$ spectrum
is studied in Figs.~\ref{fig4b},\ref{fig4d},\ref{fig4f}.
By evaluating PDF uncertainties with MSTW2008 NNLO PDFs (red band in Figs.~\ref{fig4b},\ref{fig4d},\ref{fig4f}) we see that the uncertainty is at the $\pm 1-2\%$ level, both at the Tevatron and at the LHC.
The use of different PDF sets affects not only the absolute value of the NNLO cross section (see e.g. Ref.~\cite{Watt:2011kp}) but also the shape of the $q_T$ spectrum. The predictions obtained with NNPDF 2.1 PDFs are in good agreement with those obtained with the MSTW2008 set and
the uncertainty bands overlap over a wide range of transverse momenta.
On the contrary, the prediction obtained with the ABKM09 NNLO set is softer
and the uncertainty band does not overlap with the MSTW2008 band.
This behaviour is not completely unexpected: when the Higgs boson is produced
at large transverse momenta, larger values of Bjorken $x$ are probed, where the ABKM gluon is smaller than MSTW2008 one.
The JR09 band shows a good compatibility with the MSTW2008 result, at least at the Tevatron and at the LHC for $\sqrt{s}=7$ TeV, where the uncertainty is however rather large. At the LHC for $\sqrt{s}=14$ TeV the differences with the MSTW2008 result are more pronounced.

\section{Summary}
\label{sec:summa}

In this paper we have considered the $q_T$ spectrum 
of Higgs bosons  produced in hadron collisions, and we have presented a 
perturbative QCD study
based on transverse-momentum resummation up to NNLL+NLO accuracy.

We have followed the formalism developed in 
Refs. \cite{Catani:2000vq, Bozzi:2005wk, Catani:2010pd}, which is valid for the production of a
generic high-mass system of non strongly-interacting particles 
in hadron collisions. 
The formalism combines small-$q_T$ resummation at a given logarithmic accuracy 
with the fixed-order calculations. It implements a unitarity constraint
that guarantees that the integral over $q_T$ of the differential cross section
coincides with
the total cross section at the corresponding fixed-order accuracy.
This leads to QCD
predictions with a controllable and uniform perturbative 
accuracy
over the region from small up to large values of $q_T$. 
At large values of $q_T$, the resummation formalism is superseded by customary
fixed-order calculations.

We have considered Higgs bosons produced 
 by gluon fusion in $p{\bar p}$ collisions at the Tevatron
and $pp$ collisions at LHC energies, and we have presented an update of the
phenomenological analysis of Ref.~\cite{Bozzi:2005wk}.
The calculation now includes the
exact value of the NNLO hard-collinear coefficients ${\cal H}_N^{H(2)}$ computed in Ref. \cite{Catani:2007vq,Catani:2011kr}, and the recently derived 
value of the NNLL coefficient $A^{(3)}$ \cite{Becher:2010tm}.

We have performed 
a study of the scale dependence of our results
to estimate 
the corresponding perturbative uncertainty. 
In a wide region of transverse momenta
the size of the scale uncertainties is
considerably reduced in going from
NLL+LO to NNLL+NLO accuracy.

Our calculation for the $q_T$ spectrum 
is implemented in the updated version of the numerical code {\ttfamily HqT}.
We have argued that, given the use that is currently done of our numerical program,
the important information is in the {\em shape} of the $q_T$ spectrum.
We have thus studied the uncertainties of the normalized spectrum, comparing scale
and PDF uncertainties, and estimating the impact of NP effects.

\noindent {\bf Acknowledgements.}

This work has been supported in part by the European Commission through the 'LHCPhenoNet' Initial Training Network PITN-GA-2010-264564.

\end{document}